\documentclass[aps,pra,showpacs,amsmath,amssymb,amsfonts,tightenlines,twocolumn,lengthcheck]{revtex4-1}
\usepackage{graphicx}
\graphicspath{{./}{figures/}}
\usepackage{braket}
\usepackage[a4paper,margin=1in]{geometry}
\usepackage{amsmath,amssymb,amsfonts,amsthm,mathtools,bm}
\allowdisplaybreaks[2]
\usepackage{booktabs}
\usepackage{array}
\usepackage{mathrsfs}
\usepackage[T1]{fontenc}
\usepackage[utf8]{inputenc}
\usepackage[table,xcdraw,dvipsnames]{xcolor}
\usepackage{bbm}
\usepackage{latexsym}
\usepackage{physics}
\usepackage{times,txfonts}
\DeclareFontFamily{U}{stixcalA}{\skewchar\font=45}
\DeclareFontShape{U}{stixcalA}{m}{n}{<-> stix-mathcal}{}
\DeclareFontShape{U}{stixcalA}{b}{n}{<-> stix-mathcal-bold}{}
\DeclareMathAlphabet{\stixcal}{U}{stixcalA}{m}{n}
\SetMathAlphabet{\stixcal}{bold}{U}{stixcalA}{b}{n}
\usepackage{booktabs}  
\usepackage{tabularx} 
\usepackage{makecell}
\usepackage[colorlinks=true,linkcolor=blue,citecolor=blue,urlcolor=blue]{hyperref}

\begin{document}
	\raggedbottom
	
	\title{Chirp-stabilized charging of a two-level quantum battery}
	
	\author{Yang-Yang  Yuan}
	\affiliation{Key Laboratory of Low-Dimensional Quantum Structures and Quantum Control of Ministry of Education, \\Key Laboratory for Matter Microstructure and Function of Hunan Province,\\ Department of Physics and Synergetic Innovation Center for Quantum Effects and Applications,\\ Hunan Normal University, Changsha 410081, China,\\ Hunan Research Center of the Basic Discipline for Quantum Effects and Quantum Technologies, Hunan Normal University, Changsha 410081, China,\\Institute of Interdisciplinary Studies, Hunan Normal University, Changsha 410081, China}
	
	\author{Xian-Li Yin}
	\email{Corresponding author: xianliyin@foxmail.com}
	\affiliation{Key Laboratory of Low-Dimensional Quantum Structures and Quantum Control of Ministry of Education, \\Key Laboratory for Matter Microstructure and Function of Hunan Province,\\ Department of Physics and Synergetic Innovation Center for Quantum Effects and Applications,\\ Hunan Normal University, Changsha 410081, China,\\ Hunan Research Center of the Basic Discipline for Quantum Effects and Quantum Technologies, Hunan Normal University, Changsha 410081, China,\\Institute of Interdisciplinary Studies, Hunan Normal University, Changsha 410081, China}
	
	\author{Jin-Feng Huang}
	\email{Corresponding author: jfhuang@hunnu.edu.cn}
	\affiliation{Key Laboratory of Low-Dimensional Quantum Structures and Quantum Control of Ministry of Education, \\Key Laboratory for Matter Microstructure and Function of Hunan Province,\\ Department of Physics and Synergetic Innovation Center for Quantum Effects and Applications,\\ Hunan Normal University, Changsha 410081, China,\\ Hunan Research Center of the Basic Discipline for Quantum Effects and Quantum Technologies, Hunan Normal University, Changsha 410081, China,\\Institute of Interdisciplinary Studies, Hunan Normal University, Changsha 410081, China}
	
	\author{Jie-Qiao Liao}
	\email{Corresponding author: jqliao@hunnu.edu.cn}
	\affiliation{Key Laboratory of Low-Dimensional Quantum Structures and Quantum Control of Ministry of Education, \\Key Laboratory for Matter Microstructure and Function of Hunan Province,\\ Department of Physics and Synergetic Innovation Center for Quantum Effects and Applications,\\ Hunan Normal University, Changsha 410081, China,\\ Hunan Research Center of the Basic Discipline for Quantum Effects and Quantum Technologies, Hunan Normal University, Changsha 410081, China,\\Institute of Interdisciplinary Studies, Hunan Normal University, Changsha 410081, China}

    

	\date{November 12th}


\begin{abstract}
Resonantly driven two-level quantum batteries (QBs) suffer from strong pulse-area dependence and energy backflow inherent to Rabi oscillations. To address this limitation, we propose charging of a two-level QB with chirped pulses. We investigate four charging protocols formed by pairing hyperbolic-secant (sech) or Gaussian pulse envelopes with linear or hyperbolic-tangent (tanh) frequency sweeps. We show that sufficiently strong chirp modulation converts the charging mechanism from pulse-area-governed Rabi rotation to adiabatic rapid passage. We analyze the underlying adiabatic dynamics and derive the criteria for robust chirped charging, which incorporate the sweep bandwidth, nonadiabatic transitions, and dissipative effects. Under strong chirp modulation, the final stored energy and ergotropy converge to stable plateaus that are nearly insensitive to the pulse area across an extended parameter window. Of the four schemes, the sech pulse with tanh chirp delivers the widest ergotropy plateau and the weakest pulse-area sensitivity. Compared with counterdiabatic driving, the sech–tanh protocol attains stable charging performance without demanding an extra drive quadrature. Furthermore, this chirp-enabled charging is resilient to static detuning and weak-to-moderate dissipation. This work establishes a practical route toward robust charging of quantum batteries using chirped-pulse control.
\end{abstract}

\maketitle

\section{Introduction}
\label{sec:introduction}

Quantum batteries provide a versatile framework for investigating energy storage, transfer, and release in microscopic quantum systems subject to tailored dynamical controls~\cite{PRE-042123-2013,Hovhannisyan-2013,CampaioliBook-2018,RMP-031001-2024}.
Since their inception, central figures of merit including stored energy, ergotropy, charging time, and average power have been characterized for both single-cell and charger-mediated battery architectures~\cite{PRE-042123-2013,NJP-075015-2015,PRB-205423-2018,PRL-047702-2019,PRR-023113-2020,PRL-040601-2020,PRL-130602-2022,Wen-2025-wireless,Beder-2026-cavities}.
Collective charging strategies, ranging from parallel arrangements to spin-chain and Dicke-type batteries, have been pursued to harness quantum advantages in charging performance~\cite{PRL-117702-2018,PRL-150601-2017,PRA-022106-2018,PRB-205437-2019,Ghosh-2020-interactions,Rossini-2019-MBL,Jing-2021-PRA}. While much of this work aims to boost charging metrics, coupling to the environment is inevitable in practical implementations. This has spurred intensive research on dissipative effects in quantum-battery charging~\cite{NJP-083085-2020,PRA-033715-2021,PRA-023725-2023,PRA-032211-2024,PRL-130401-2025}, non-Markovian charging dynamics~\cite{NJP-083007-2020,QST-035007-2023,PRA-032202-2026}, and self-discharge and aging~\cite{Pirmoradian-2019-aging,Santos-2021-self-discharge,PRL-220402-2025,PRL-020405-2025}. Complementary approaches including dissipative stabilization~\cite{PRL-210601-2019,PRApplied-024092-2020,PRL-210402-2024,PRA-012617-2026,Guo-2026-waveguide}, reservoir-assisted charging~\cite{PRA-060201-2020,PRL-090401-2024,Lu-2025-topological,Yan-2026-giant-atom}, and measurement-feedback open-system control~\cite{PRR-013095-2020,Quantum-500-2021,PRE-014138-2022,PRR-L032035-2026,PRL-050406-2026} have been developed to enhance stored energy and ergotropy. At the same time, experimental demonstrations in organic microcavities and superconducting circuits~\cite{SCA-2022,QST-045018-2022,PRR-023091-2024,PRL-060401-2026} show that quantum-battery research is advancing toward experimentally feasible protocols built upon high-precision quantum control~\cite{PRA-042601-2022,PRA-L03201-2023}.

Among available control schemes, pulse-shaped charging stands out, as the amplitude, duration, phase, and frequency of laser or microwave drives can be precisely engineered. Such pulse control has been explored for driven two-level quantum batteries~\cite{PRE-052106-2019,NJP-063057-2020}, optimal-control charging~\cite{PRA-032218-2023, Evangelakos-2024-optimal}, frequency-modulated superconducting qutrit batteries~\cite{APL-154002-2023}, pulsed charger–battery setups~\cite{EPL-10001-2024}, and shortcut-to-adiabaticity-assisted charging~\cite{FrontPhys-31503-2022,PLA-130124-2025}. Pulse-based charging is inherently a finite-time transient process, where coherent Rabi oscillations dominate energy exchange~\cite{NJP-063057-2020,EPL-10001-2024}. In open quantum batteries, relaxation and dephasing introduce competing dynamical time scales~\cite{PRB-035421-2019,PRL-210601-2019,PRR-013095-2020,NJP-083007-2020,NJP-083085-2020}. For unchirped resonant pulses, population transfer relies on standard Rabi rotation~\cite{NJP-063057-2020,LJ-1975}.
Though fast, this mechanism hinges on an accurate pulse area and perfect resonance. Fluctuations in peak amplitude, interaction time, or static detuning easily degrade population transfer: insufficient rotation yields partial charging, while over-rotation pumps population back to the ground state and causes energy backflow. Robust, fast pulse charging therefore remains an outstanding challenge.

Frequency chirping across resonance offers an alternative population-transfer pathway. Under adiabatic conditions, the sweep implements adiabatic rapid passage for the two-level system, enabling the population to follow an instantaneous dressed eigenbranch rather than requiring a carefully tuned resonant Rabi pulse~\cite{ARP-2001,PRA-023409-2002,PRA-063422-2007,PRA-033411-2011}. This transfer is known to be resilient against variations in pulse amplitude, timing, and detuning~\cite{PRA-023409-2002,PRA-063422-2007}, and adiabatic driving has been shown to stabilize quantum-battery charging against environmental noise~\cite{FrontPhys-31503-2022,PRE-032107-2019}. Still, for dissipative quantum batteries, it is unclear whether chirp-enabled adiabatic passage can suppress back-transfer and stabilize stored energy and ergotropy, and how this performance depends on pulse envelope, chirp profile, and dissipation strength.

In this work, we investigate charging of a two-level quantum battery under four representative pulse–chirp combinations: sech or Gaussian envelopes paired with linear or tanh frequency sweeps. Open-system dynamics are modeled by a local Gorini–Kossakowski–Sudarshan–Lindblad (GKSL) master equation that includes relaxation, thermal excitation, and pure dephasing~\cite{GKSL-1976,Lindblad-1976,BreuerPetruccione-2002}. We characterize the crossover from pulse-area-governed Rabi charging to chirp-mediated adiabatic rapid passage via the dressed-state energy gap, a local adiabaticity parameter, and dressed-state following fidelity. By combining local Landau–Zener theory with global sweep-range and dissipative bounds, we derive approximate conditions for robust adiabatic charging and compare the predicted adiabatic boundaries with numerical phase diagrams. We then quantify stored energy, ergotropy, ergotropy-to-energy ratio, and average charging power, disentangling the separate roles of pulse shape and frequency modulation. The chirped protocols are further compared with counterdiabatic driving and tested against static detuning and weak-to-moderate dissipation.
Lastly, we discuss experimentally viable parameters for superconducting-circuit platforms and establish global criteria on sweep coverage and dissipation for stable adiabatic charging.

The remainder of the paper is organized as follows. In Sec.~\ref{sec:model}, we introduce the driven two-level battery model, define the chirped-pulse protocols, and present the open-system dynamics governed by the local GKSL master equation.
In Sec.~\ref{Adiabaticity Diagnostics and Pulse Design}, we introduce adiabaticity diagnostics, derive Landau–Zener-based estimates for nonadiabatic transitions and the corresponding chirp thresholds, and map out adiabatic operating regions.
In Sec.~\ref{Chirped-stabilized charging performance}, we analyze the chirp-stabilized charging performance and compare it with that of a counterdiabatic STA protocol. In Sec.~\ref{Discussion}, we address experimental feasibility and state the global sweep-coverage and dissipative criteria for robust adiabatic passage. We conclude in Sec.~\ref{conclusion}.

\section{Physical Model and Theoretical Framework}
\label{sec:model}
In this section, we introduce the driven two-level quantum battery model. Specifically, we derive the rotating-frame Hamiltonian, present four different chirped-pulse protocols, and describe the open-system dynamics via a local GKSL master equation.
\subsection{Two-level quantum battery and charging interaction}
\label{sec:driven_battery}
\begin{figure}[t]
	\centering
	\includegraphics[width=0.47\textwidth,height=1.2\textheight,keepaspectratio]{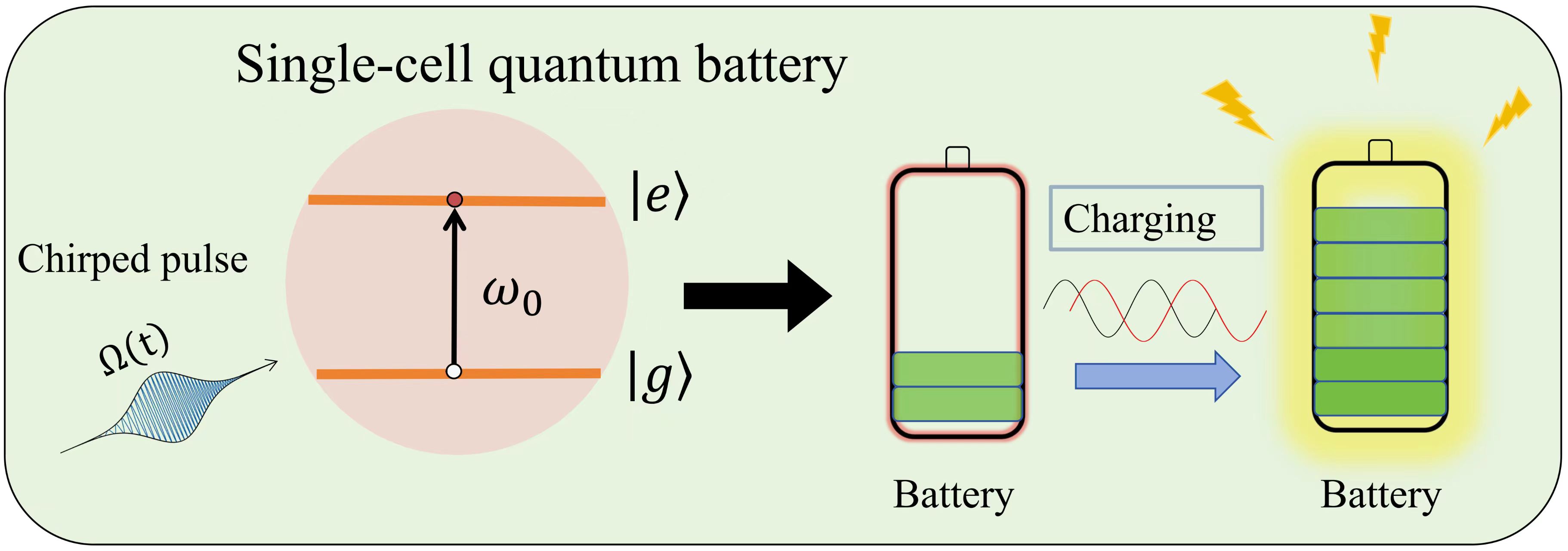}
	\caption{Schematic of a single-cell quantum battery charged by a chirped pulse. The battery cell is modeled as a two-level system with ground state $|g\rangle$, excited state $|e\rangle$, and transition frequency $\omega_0$. The chirped pulse transfers population from $|g\rangle$ to $|e\rangle$, thereby injecting energy into the cell. The right panel illustrates the corresponding charging process from a low-charge state to a high-charge state.}
	\label{fig:model}
\end{figure}

We consider a single battery cell driven by an external classical chirped pulse. The cell is modeled as a two-level system with ground state $|g\rangle$ and excited state $|e\rangle$, separated by a transition frequency $\omega_0$, as shown in Fig.~\ref{fig:model}. The battery is initially prepared in $|g\rangle$, and the charging process transfers the population from $|g\rangle$ toward $|e\rangle$. The total Hamiltonian of the driven system reads
\begin{equation}
H_s(t)=H_b+W(t)H_{\rm ch}(t),
\label{eq:total_hamiltonian}
\end{equation}
where $H_b$ is the bare Hamiltonian of the battery and $H_{\rm ch}(t)$ is the charging-interaction Hamiltonian. The switching function $W(t)$ restricts the dynamics to the charging interval,
\begin{equation}
W(t)=\Theta(t-t_i)-\Theta(t-t_f),
\label{eq:charging_window}
\end{equation}
where $\Theta(\cdot)$ denotes the Heaviside step function, and $t_i$ and $t_f$ are the initial and final times of the charging window, respectively.

In the bare-state basis $\{|e\rangle=(1,0)^T,|g\rangle=(0,1)^T\}$, the battery and charging-interaction Hamiltonians are, respectively, given by
\begin{subequations}
\label{eq:bare_charging_hamiltonian}
\begin{align}
H_b&=\frac{\hbar\omega_0}{2}\sigma_z, \\
H_{\rm ch}(t)&=\hbar\Omega(t)\cos\left[\phi_0+\int_{0}^{t}\omega_L(t')dt'\right]\sigma_x .
\end{align}
\end{subequations}
Here, $\sigma_z=|e\rangle\langle e|-|g\rangle\langle g|$ and $\sigma_x=|g\rangle\langle e|+|e\rangle\langle g|$. The quantities $\Omega(t)$, $\phi_0$, and $\omega_L(t)$ are the time-dependent Rabi-frequency envelope, initial phase, and instantaneous driving frequency, respectively.
\subsection{The Hamiltonian in the rotating frame and different chirped-pulse protocols}
\label{sec:rotating_frame_protocols}

We define the pulse phase $\phi(t)=\phi_0+\int_0^t\omega_L(t')dt'$ and transform to a rotating frame via $U(t)=\exp[i\phi(t)\sigma_z/2]$. Under the rotating-wave approximation, the effective Hamiltonian is
\begin{equation}
H_{\rm eff}(t)=\frac{\hbar}{2}\left[W(t)\Omega(t)\sigma_x+\delta(t)\sigma_z\right],
\label{eq:heff}
\end{equation}
with $\delta(t)=\omega_0-\omega_L(t)$ the instantaneous detuning. Equation~\eqref{eq:heff} is equivalent to the Hamiltonian of a spin-$1/2$ particle subject to an effective field $\mathbf{B}_{\rm eff}(t)=(\Omega(t),0,\delta(t))$. The transverse component $\Omega(t)$ induces transitions between $|g\rangle$ and $|e\rangle$, whereas the longitudinal component $\delta(t)$ controls the instantaneous energy splitting in the rotating frame.
\begin{figure}[tbp]
\centering
\includegraphics[width=0.85\columnwidth]{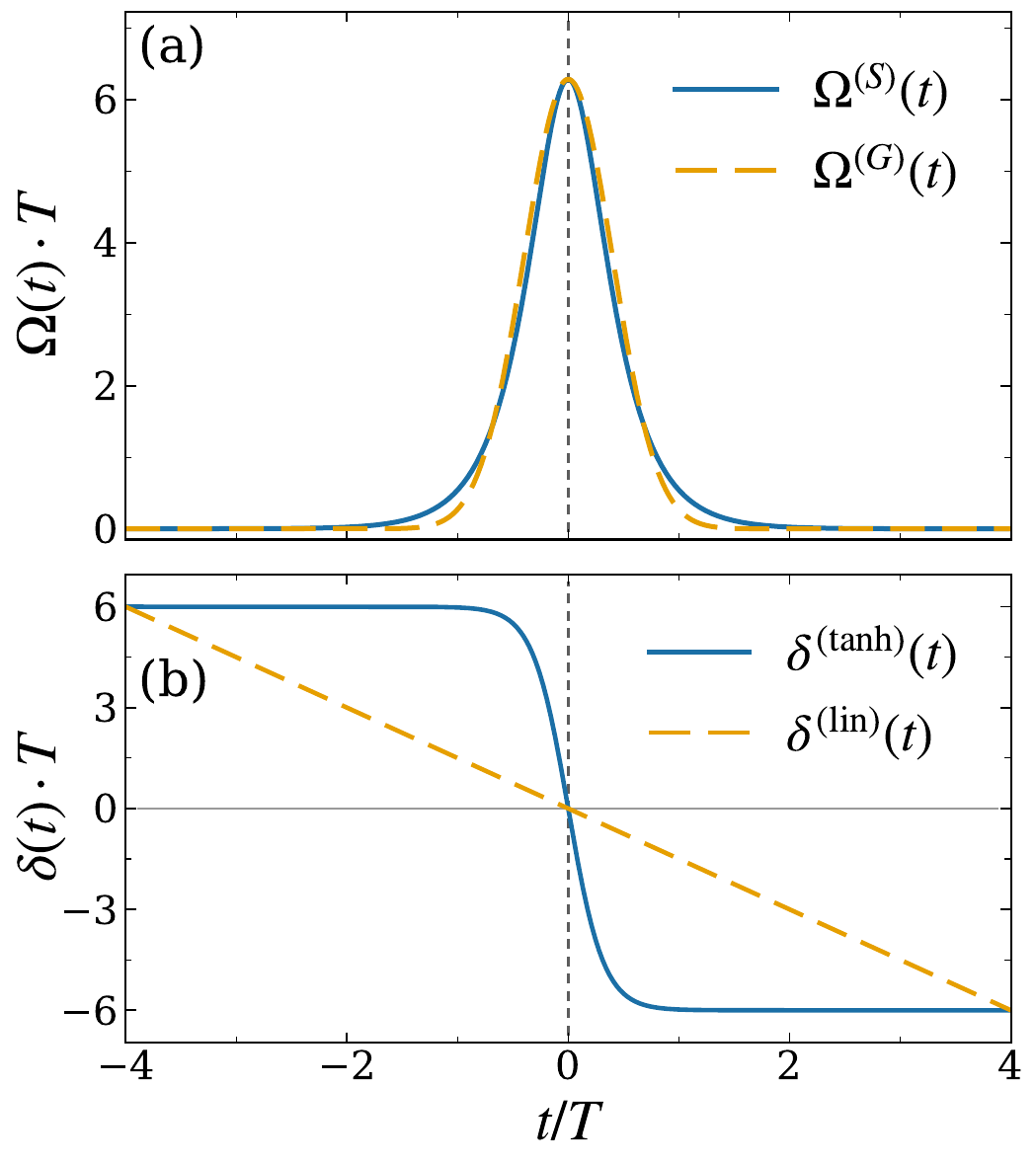}
\caption{Pulse envelopes and detuning protocols. (a) Normalized Rabi-frequency envelopes $\Omega^{(S)}(t) \cdot T$ and $\Omega^{(G)}(t) \cdot T$ as functions of $t/T$ for the sech (solid curve) and Gaussian (dashed curve) pulses, respectively. Both pulses are centered at $t=0$, as marked by the vertical dashed line. (b) Detunings $\delta^{(\rm lin)}(t) \cdot T$ (dashed line) and
$\delta^{(\rm tanh)}(t)\cdot T$ (solid line) versus $t/T$ for the linear and tanh chirps, respectively. The dimensionless parameters used are $T=\omega_0^{-1}$, $A=\pi$, $\Delta=0$, $\alpha=1.5\omega_0^2$, and $\delta_0=6\omega_0$.}
\label{Fig-Graph2}
\end{figure}
To systematically investigate the effects of pulse shaping and frequency modulation, we consider two representative pulse envelopes: a sech pulse and a Gaussian pulse,
\begin{subequations}
\label{eq:pulse_envelopes}
\begin{align}
\Omega^{(S)}(t)&=\frac{2A}{T}\operatorname{sech}\left(\frac{\pi t}{T}\right), \\
\Omega^{(G)}(t)&=\frac{2A}{T}\exp\left[-\pi\left(\frac{t}{T}\right)^2\right],
\end{align}
\end{subequations}
where the parameter $T$ sets the temporal width, and the dimensionless parameter $A$ controls the overall pulse strength. The superscripts (S) and (G) in Eqs.~(\ref{eq:pulse_envelopes}) label the two pulse shapes.
The common prefactor $2A/T$ and the time scalings are chosen to equalize the integrated areas of the two envelopes. Thus, both envelopes share the same pulse area
\begin{equation}
\stixcal{A}\equiv\int_{-\infty}^{+\infty}\Omega(t)dt=2A,
\label{eq:pulse_area}
\end{equation}
which enables a fair comparison of their charging performance at equal pulse areas. As shown in Fig.~\ref{Fig-Graph2}(a), the sech pulse (blue solid curve) features longer temporal tails, whereas the Gaussian pulse (yellow dashed curve) is more strongly localized around the pulse center.

For frequency modulation, we adopt linear and tanh chirps. The corresponding instantaneous driving frequencies read
\begin{subequations}
\label{eq:driving_frequency}
\begin{align}
\omega_L^{(\rm lin)}(t)&=\omega_{L0}+\alpha t,\\
\omega_L^{(\rm tanh)}(t)&=\omega_{L0}+\delta_0\tanh\left(\frac{\pi t}{T}\right),
\end{align}
\end{subequations}
where $\alpha$ is the linear chirp rate and $\delta_0$ denotes the modulation amplitude of the tanh chirp, with $\alpha>0$ and $\delta_0>0$. The associated detuning functions are
\begin{subequations}
\label{eq:detuning_protocols}
\begin{align}
\delta^{(\rm lin)}(t)&=\Delta-\alpha t,\\
\delta^{(\rm tanh)}(t)&=\Delta-\delta_0\tanh\left(\frac{\pi t}{T}\right).
\end{align}
\end{subequations}
Here, $\Delta=\omega_{0}-\omega_{L0}$ is the detuning at the pulse center.
Figure~\ref{Fig-Graph2}(b) displays the detuning sweeps for these two protocols. The linear chirp (yellow dashed curve) scans the detuning at a constant rate $\alpha$, while the tanh chirp (blue solid curve) implements a bounded, smooth frequency sweep, with the fastest variation concentrated near the pulse center at $t_{\ast}=0$. The tanh chirp asymptotically approaches $\Delta\pm \delta_{0}$, yielding a finite detuning window tunable via $\delta_0$.

The sech envelope paired with a tanh detuning sweep is closely related to the Allen--Eberly/Hioe~\cite{LJ-1975,Hioe-1984} and Demkov--Kunike~\cite{Suominen-1992,Torosov-2011} families of analytically tractable two-state control models. In these models, sech-shaped couplings and tanh-shaped level crossings are standard building blocks for population inversion and adiabatic passage.
The resonance condition $\delta(t_{\ast})=0$ yields the crossing time
\begin{equation}
	t_{\ast}=\begin{cases}
		\dfrac{\Delta}{\alpha}, & \alpha\neq 0,\\[1.0em]
		\dfrac{T}{\pi}\operatorname{artanh}\left(\dfrac{\Delta}{\delta_0}\right), & |\Delta|<|\delta_0|,
	\end{cases}
	\label{eq:resonance_times}
\end{equation}
where the two entries correspond to the linear and tanh chirps, respectively.
The condition $|\Delta|<|\delta_0|$ guarantees that the tanh sweep crosses resonance. In the symmetric case $\Delta=0$, resonance occurs exactly at the pulse center $t_{\ast}=0$.

\subsection{Local GKSL master equation}
\label{Local GKSL master equation}
In a realistic setting, the two-level quantum battery unavoidably couples to its environment. The resulting charging dynamics is captured by a local Markovian GKSL master equation written in the bare-state basis~\cite{GKSL-1976,Lindblad-1976,BreuerPetruccione-2002,Davies-1974},
\begin{equation}
	\label{eq:gksl_master}
	\begin{aligned}
		\dot{\rho}_s(t)=&-\frac{i}{\hbar}[H_{\rm eff}(t),\rho_s(t)]
		+\Gamma_{\downarrow}\mathcal{D}[\sigma_-]\rho_s(t)\\
		&+\Gamma_{\uparrow}\mathcal{D}[\sigma_+]\rho_s(t)
		+\frac{\gamma_\phi}{2}\mathcal{D}[\sigma_z]\rho_s(t),
	\end{aligned}
\end{equation}
where $\rho_s(t)$ denotes the reduced density operator of the battery. The rates $\Gamma_{\downarrow}$, $\Gamma_{\uparrow}$, and $\gamma_\phi$ correspond to spontaneous relaxation, incoherent excitation, and pure dephasing, respectively. Here, $\sigma_+=|e\rangle\langle g|$ and $\sigma_-=|g\rangle\langle e|$ are the raising and lowering operators. The superoperator in Eq.~(\ref{eq:gksl_master}) is defined as
$\mathcal{D}[o]\rho_s=o\rho_so^\dagger-\{o^\dagger o,\rho_s\}/2$.
We adopt Eq.~\eqref{eq:gksl_master} as an effective Markovian description valid in the weak-dissipation regime. For sufficiently strong or rapidly modulated driving, this local bare-basis formulation may break down, and a dressed-basis, global, or explicitly time-dependent nonadiabatic master-equation treatment can become necessary~\cite{LevyKosloff-2014,Hofer-2018,Dann-2018}.

For a weakly coupled thermal bosonic bath, the excitation and relaxation rates at the transition frequency $\omega_0$ read
\begin{subequations}
	\label{eq:gamma}
	\begin{align}
		\Gamma_{\uparrow}&=2\pi J(\omega_0)n(\omega_0),\\
		\Gamma_{\downarrow}&=2\pi J(\omega_0)[n(\omega_0)+1],			
	\end{align}
\end{subequations}
where $n(\omega)=[\exp(\hbar\omega/k_BT_{\rm bath})-1]^{-1}$ is the Bose--Einstein distribution function. For an Ohmic bath, we use $J(\omega)=2\kappa\omega e^{-\omega/\omega_c}$, with $\kappa$ the dimensionless coupling strength and $\omega_c$ the cutoff frequency.

In the ordered bare-state basis $\{|e\rangle,|g\rangle\}$, the density matrix takes the form
\begin{equation}
\label{eq:density_matrix}
	\rho_s(t)=\begin{pmatrix}
		\rho_{ee}(t) & \rho_{eg}(t)\\
		\rho_{ge}(t) & \rho_{gg}(t)
	\end{pmatrix}.
\end{equation}
Subject to the initial condition $\rho_s(t_i)=|g\rangle\langle g|$, Eq.~\eqref{eq:gksl_master} yields the equations of motion for the excited-state population and the off-diagonal coherence 
	\begin{align}
    \label{eq:rho_eee}
	 		\dot{\rho}_{ee}(t)
		&=-\frac{iW(t)}{2}\Omega(t)
		[\rho_{ge}(t)-\rho_{eg}(t)]\\\notag
		   &\quad-\Gamma_{\downarrow}\rho_{ee}(t)
		+\Gamma_{\uparrow}\rho_{gg}(t),\\
        \label{eq:density_matrix_equations}
		\dot{\rho}_{eg}(t)
		&=+\frac{iW(t)}{2}\Omega(t)
		\left[\rho_{ee}(t)-\rho_{gg}(t)\right]\\\notag
        &\quad -i\delta(t)\rho_{eg}(t)-\Gamma_2\rho_{eg}(t).
	\end{align}
Here $\rho_{gg}(t)=1-\rho_{ee}(t)$ and $\rho_{ge}(t)=\rho_{eg}^*(t)$. In Eq.~(\ref{eq:density_matrix_equations}), $\Gamma_2=(\Gamma_{\uparrow}+\Gamma_{\downarrow})/2+\gamma_\phi$ is the transverse decoherence rate. 
{For the Hamiltonian in Eq.~(\ref{eq:heff}), the charging dynamics is obtained by integrating Eqs.~(\ref{eq:rho_eee}) and~(\ref{eq:density_matrix_equations}) over the time window $[t_i,t_f]=[-4T,4T]$.

\section{Adiabaticity Diagnostics and Pulse Design}
\label{Adiabaticity Diagnostics and Pulse Design}
In this section, we analyze chirped-pulse charging in the instantaneous dressed-state basis and introduce the quantities used to assess the adiabaticity. Based on these quantities, we derive the local Landau–Zener criterion and analyze the corresponding adiabatic operating regime.

\subsection{Adiabatic representation and diagnostic quantities}
\label{sec:adiabatic_representation}

The population transfer induced by the frequency sweep can be analyzed in the instantaneous eigenstates of the effective Hamiltonian $H_{\rm eff}(t)$ in Eq.~(\ref{eq:heff}). Diagonalizing $H_{\rm eff}(t)$ gives the dressed states
\begin{subequations}
\begin{align}
|\chi_-(t)\rangle&=\cos[\theta(t)/2]|g\rangle-
\sin[\theta(t)/2]|e\rangle,\\
|\chi_+(t)\rangle&=\sin[\theta(t)/2]|g\rangle+
\cos[\theta(t)/2]|e\rangle,
\label{eq:instantaneous_eigenstates+}
\end{align}
\end{subequations}
which satisfy $H_{\rm eff}(t)|\chi_\pm(t)\rangle=E_\pm(t)|\chi_\pm(t)\rangle$, with $E_\pm(t)=\pm\hbar\Upsilon(t)/2$. Here the effective Rabi frequency $\Upsilon(t)$ and the mixing angle $\theta(t)$ are, respectively, defined as
\begin{subequations}
\label{eq:mixing_angle}
\begin{align}
\Upsilon(t)&=\sqrt{\Omega^2(t)+\delta^2(t)},\\
\cos\theta(t)&=\frac{\delta(t)}{\Upsilon(t)},\qquad
\sin\theta(t)=\frac{\Omega(t)}{\Upsilon(t)}.
\end{align}
\end{subequations}

The population transfer during the frequency sweep can be understood from the evolution of the dressed-state branch $|\chi_-(t)\rangle$.
At the beginning of the sweep, $\delta(t)>0$ and $|\delta(t)|\gg\Omega(t)$. In this limit, $\theta(t)\simeq0$ and $|\chi_-(t)\rangle\simeq|g\rangle$. As the drive passes through resonance, $\theta(t)$ changes continuously and the dressed state evolves from a ground-state-like state. After the crossing, $\delta(t)<0$ and $|\delta(t)|\gg\Omega(t)$. In this regime, $\theta(t)\simeq \pi$ and $|\chi_-(t)\rangle \simeq- |e\rangle $. If the transitions between the two dressed-state branches remain negligible, the system follows $|\chi_-(t)\rangle$ throughout the sweep and the population is transferred from $|g\rangle $ to $|e\rangle $ through adiabatic rapid passage.

To quantify departures from adiabatic following, we transform the Hamiltonian into the instantaneous eigenbasis representation. Applying the unitary rotation $U(t)=\exp[i\theta(t)\sigma_y/2]$ to $H_{\rm eff}(t)$ in Eq.~\eqref{eq:heff} gives
\begin{equation}
H_{\rm ad}(t)=\frac{\hbar\Upsilon(t)}{2}\sigma_z-
\frac{\hbar\dot{\theta}(t)}{2}\sigma_y .
\label{eq:Had}
\end{equation}
The first term in Eq.~\eqref{eq:Had} is the instantaneous dressed-state splitting, while the second term is the nonadiabatic coupling generated by the time dependence of the basis.
The nonadiabatic term in Eq.~\eqref{eq:Had} couples the instantaneous dressed states separated by an energy gap $\hbar\Upsilon(t)$. For the Hamiltonian evolution, a commonly used local adiabaticity criterion, motivated by the quantum adiabatic theorem \cite{Adiabatensatzes,TKato}, is given by
\begin{equation}
	\frac{\hbar\left|\langle\chi_{+}(t)|
		\dot{H}_{\rm eff}(t)|\chi_{-}(t)\rangle\right|}
	{|E_+(t)-E_-(t)|^2}\ll1.
	\label{eq}
\end{equation}
Using $\langle\chi_{+}(t)|\dot H_{\rm eff}(t)|\chi_{-}(t)\rangle=\hbar\Upsilon\dot\theta(t)/2$ and $\dot\theta(t)=[\delta(t)\dot\Omega(t)-\Omega(t)\dot\delta(t)]/\Upsilon^2(t)$, this criterion yields
\begin{equation}
	\Lambda(t)=\frac{|\dot\theta(t)|}{2\Upsilon(t)}
	=\frac{|\delta(t)\dot\Omega(t)-\Omega(t)\dot\delta(t)|}
	{2\left[\Omega^2(t)+\delta^2(t)\right]^{3/2}}\ll1.
	\label{eq:local_adiabaticity}
\end{equation}
Thus, $\Lambda(t)$ quantifies the nonadiabatic coupling relative to the instantaneous energy gap. It characterizes coherent nonadiabatic transitions, while relaxation and dephasing are treated separately through the GKSL master equation.
\begin{figure*}[htbp]
	\centering
	\includegraphics[width=\textwidth]{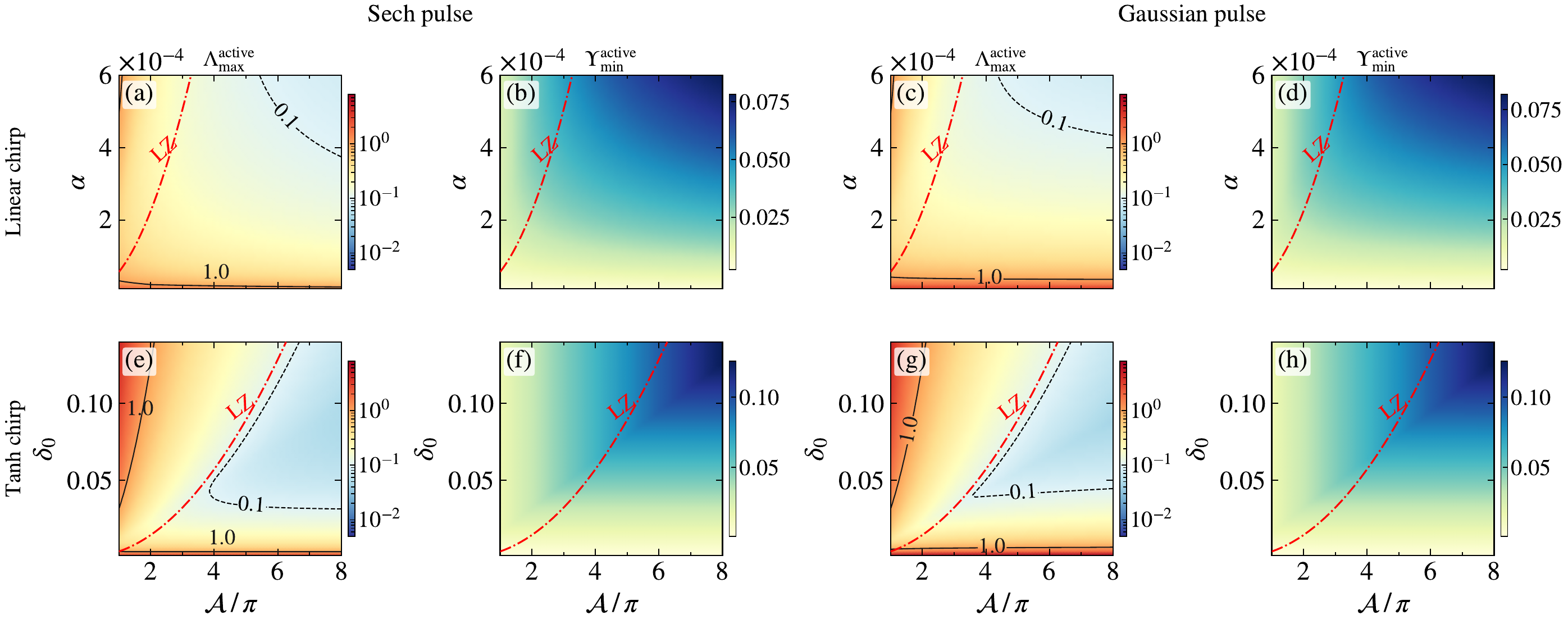}
	\caption{Adiabaticity phase diagrams for the sech and Gaussian pulses. The upper and lower rows correspond to the linear and tanh chirps, respectively. The black contour lines correspond to $ \Lambda_{\max}^{\mathrm{active}}=0.1$ and 1.0, respectively. The former serves as a practical benchmark for adiabatic evolution, whereas the latter indicates that the nonadiabatic coupling becomes comparable to the instantaneous dressed-state splitting. (a,~e) and (c,~g) Active-window maximum $\Lambda_{\max}^{\mathrm{active}}$ with the Landau–Zener reference curves (red dot-dashed). (b,~f) and (d,~h) Active-window minimum dressed-state gap $\Upsilon_{\min}^{\mathrm{active}}$. The parameters used are $T=200\omega_0^{-1}$, $\Delta=0$, and $\epsilon=10^{-3}$.}
	\label{fig:sech}
\end{figure*}

To compare the chirped protocols, we evaluate the adiabaticity within the active pulse window
\begin{equation}
	I_{\rm active}=
	\{t\in[t_i,t_f]\mid\Omega(t)>\epsilon\Omega_{\max}\},
	\quad \epsilon=10^{-3},
	\label{eq:active_window}
\end{equation}
where $\epsilon$ is a numerical cutoff excluding negligible pulse tails. We characterize this window by $\Lambda_{\max}^{\rm active}=\max_{t\in I_{\rm active}}\Lambda(t)$ and $\Upsilon_{\min}^{\rm active}=\min_{t\in I_{\rm active}}\Upsilon(t)$, which quantify the maximum relative nonadiabatic coupling and the minimum dressed-state gap, respectively. These quantities characterize the coherent adiabatic regime, while complete population inversion additionally requires sufficient sweep coverage. For a more detailed discussion, see Sec.~\ref{app:validity} below.

\subsection{Landau–Zener criterion and chirp boundaries}
\label{sec:analytical_criterion}

To obtain an analytical estimate of the nonadiabatic transition probability, we focus on the resonance-crossing time \(t_{\ast}\), defined by \(\delta(t_{\ast})=0\), in the regime of sufficiently strong chirping. Near the crossing, the detuning can be linearized and the pulse envelope can be approximated by its local value,
\begin{subequations}
	\begin{align}
		\delta(t)&\simeq\dot{\delta}(t_{\ast})(t-t_{\ast})=-v_*(t-t_{\ast}),\\
		\Omega(t)&\simeq\Omega(t_{\ast})=\Omega_*,
		\label{eq:local_lz_approx}
	\end{align}
\end{subequations}
where $v_*=|\dot{\delta}(t_{\ast})|$ is the magnitude of the local sweep rate. At the resonance crossing, the dressed-state gap is $\Upsilon(t_{\ast})=\Omega_*$. This value equals the active-window minimum gap $\Upsilon_{\min}^{\rm active}$ only when the minimum gap occurs at $t=t_{\ast}$. Otherwise, the minimum gap occurs away from resonance.

For a locally linear detuning and an approximately constant coupling, the nonadiabatic transition probability is estimated by the Landau–Zener formula~\cite{PRA-063422-2007,PRA-033411-2011,Landau-1932,Zener-1932}
\begin{subequations}
	\begin{align}
		P_{\rm na}&\simeq\exp\left(-\frac{\pi\Omega_*^2}{2v_*}\right)
		=\exp\left(-\frac{\pi}{4\Lambda_*}\right),\label{eq:lz_probability}
		\\
		\Lambda_* &\simeq\frac{v_*}{2\Omega_*^2}.\label{eq:lz_probability.}
	\end{align}
\end{subequations}
The characteristic time scale for the avoided crossing is $\tau_{\rm LZ}\sim\Omega_*/v_*$. The Landau–Zener approximation requires the pulse envelope to vary little and the detuning to remain approximately linear over this interval. For $\Delta=0$, the crossing occurs at $t_{\ast}=0$, where $\dot{\Omega}(0)=0$ for both envelopes. Moreover, the linear chirp is exactly linear in time, while the curvature of the tanh sweep vanishes at the pulse center. Therefore, the approximation is well suited to the central resonance crossing. For $\Delta\neq0$, the crossing shifts away from the pulse center, and Eq.~(\ref{eq:lz_probability}) should be regarded as a local estimate of the nonadiabatic leakage.

For a target nonadiabatic error $\eta_{\rm na}=10^{-3}$, the condition $P_{\rm na}\leq\eta_{\rm na}$ gives
\begin{equation}
	\Lambda_{\ast} \leq \frac{\pi}{4\ln(1/\eta_{\rm na})}.
	\label{eq:alpha_bound}
\end{equation}
Thus, the numerical threshold $\Lambda_{\max}^{\rm active}=0.1$ is close to the local threshold value $\Lambda_{\ast}\simeq0.114$,
while imposing a slightly smaller upper bound on the adiabaticity parameter. For the resonant case $\Delta=0$, one has $t_{\ast}=0$ and $\Omega_*=\stixcal{A}/T$ for both pulse envelopes. The sweep rates for the linear and tanh chirps are, respectively, given by
\begin{equation}
\label{eq:sweep_rates}
	v_{\rm lin}=\alpha,
	\qquad
	v_{\rm tanh}=\frac{\pi\delta_0}{T}.
\end{equation}
According to Eq.~(\ref{eq:lz_probability}), the target condition $P_{\rm na} \leq \eta_{\rm na}$ imposes the following upper bound on the sweep rate $v_{\ast}\leq\pi\Omega^{2}_{\ast}/2\ln(1/\eta_{\rm na}) $. Substituting $\Omega_{\ast}=\stixcal{A}/T$ and the two sweep rates from Eq.~(\ref{eq:sweep_rates}) gives
\begin{subequations}
	\begin{align}
		\alpha&\lesssim \frac{\pi\stixcal{A}^2}{2T^2\ln(1/\eta_{\rm na})},\label{eq:alpha_bound}\\
		\delta_0&\lesssim \frac{\stixcal{A}^2}{2T\ln(1/\eta_{\rm na})} .\label{eq:delta0_bound}
	\end{align}
\end{subequations}
These inequalities set the upper bounds on the chirp parameters required to maintain the nonadiabatic transition probability below $\eta_{\rm na}$.
Faster sweeps increase the leakage near the resonance crossing.

\subsection{Adiabatic operating regimes}
\label{sec:adiabaticity_diagnostics}

For each protocol, $\Lambda_{\max}^{\rm active}$ characterizes the maximum nonadiabatic coupling relative to the instantaneous dressed-state gap, while $\Upsilon_{\min}^{\rm active}$ gives the minimum gap within the active pulse window. In Figs.~\ref{fig:sech}[(a), (c), (e), (g)], the black contours at $\Lambda_{\max}^{\rm active}=0.1$ and 1.0 distinguish different adiabaticity regimes. The former limits the coupling-to-gap ratio to 0.1, yielding $ P_{\rm na}\lesssim3.9\times10^{-4}$ within the local Landau–Zener approximation, whereas the latter marks where the local adiabaticity condition fails.
\begin{figure*}[t]
	\centering
	\includegraphics[width=\textwidth]{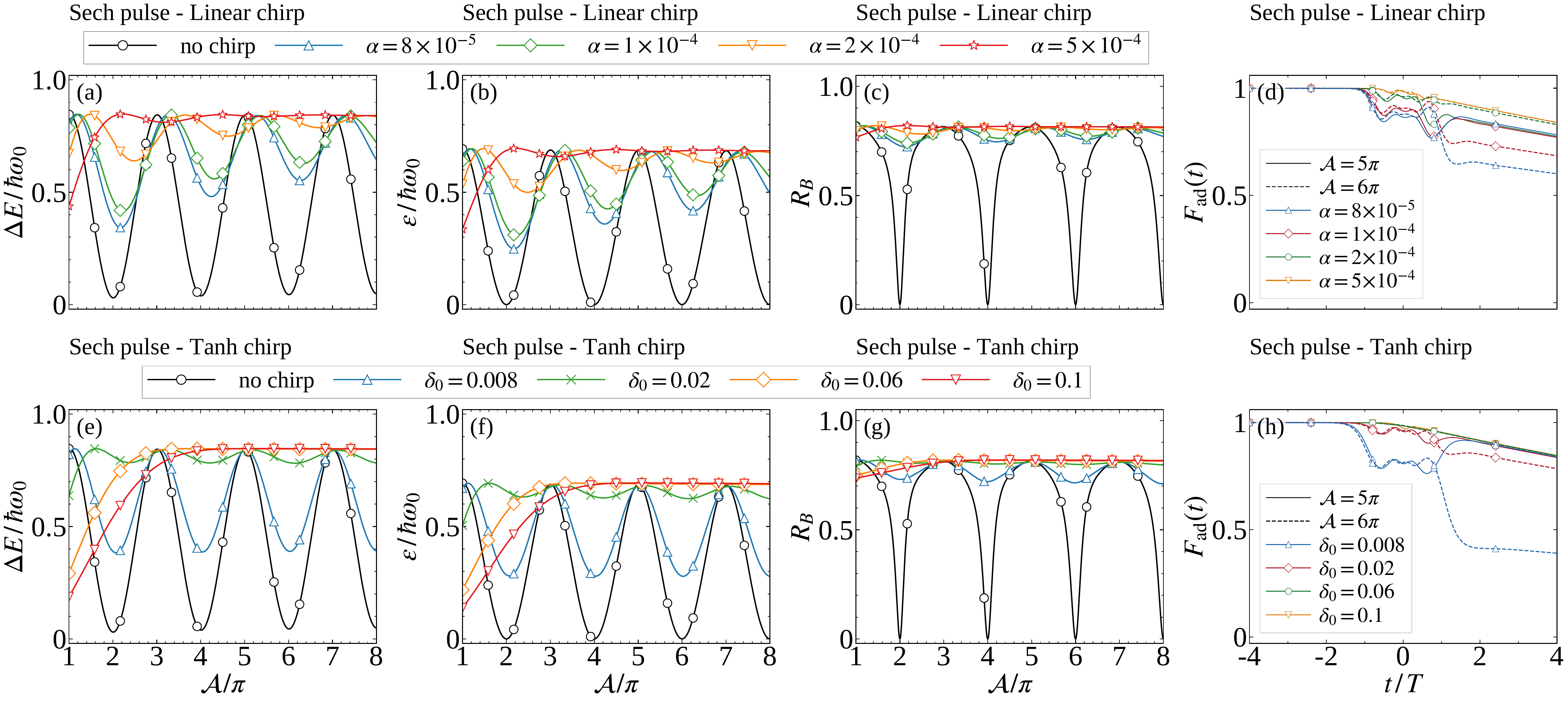}
	\caption{(a)--(c) Charging performance of a sech-driven two-level battery. The stored energy $\Delta E/\hbar\omega_0$, ergotropy $\varepsilon/\hbar\omega_0$, and ergotropy-to-energy ratio $R_{B}$ versus the pulse area $\stixcal{A}/\pi$ for the linear chirp $\delta^{\rm lin}(t)$. (e)--(g) The same quantities for the tanh chirp $\delta^{\rm tanh}(t)$.
		The unchirped case is included as a reference and exhibits pulse-area-sensitive Rabi oscillations.
		Near-zero-energy points with $\Delta E/(\hbar\omega_0)\leq10^{-6}$ are omitted from the $R_B$ curves. (d,~h) Adiabatic-following fidelity $F_{\mathrm{ad}}(t)$ for selected chirp strengths at $\stixcal{A}=5\pi$~(solid) and
			$\stixcal{A}=6\pi$~(dashed).
		Other parameters used are $T=200\omega_0^{-1}$, $\Delta=0$, $\Gamma_\downarrow=2.0\times10^{-4}\omega_0$, $\gamma_\phi=2.0\times10^{-4}\omega_0$, $\Gamma_\uparrow=\Gamma_\downarrow\exp(-\beta\hbar\omega_0)$, and $\beta=5(\hbar\omega_0)^{-1}$.}
	\label{fig:sech_triptych}
\end{figure*}

For the linear chirp [Figs.~\ref{fig:sech}(a)--\ref{fig:sech}(d)], the resonance-crossing parameter is $\Lambda_*\simeq\alpha T^2/2\stixcal{A}^2$. Thus, increasing the pulse area improves local adiabaticity, whereas increasing $\alpha $ enhances nonadiabatic transitions near resonance. For the tanh chirp [Figs.~\ref{fig:sech}(e)--\ref{fig:sech}(h)], $\Lambda_*\simeq\pi\delta_0T/2\stixcal{A}^2$, indicating that increasing $\delta_0 $ improves sweep coverage but also enhances nonadiabatic coupling near resonance. The sech and Gaussian envelopes differ mainly in their off-resonant dynamics and minimum dressed-state gaps. 
{The gap maps in Figs.~\ref{fig:sech}[(b),~(d),~(f),~(h)] show that increasing the pulse area or chirp strength generally enlarges the minimum dressed-state gap $\Upsilon_{\min}^{\mathrm{active}}$ . When the nonadiabatic coupling remains small relative to this gap, the system follows the dressed-state branch from $|g\rangle$ to $|e\rangle$, facilitating population transfer. For the tanh sweep, the bounded detuning sets the minimum gap near $\delta_{0}$ when it occurs in the weak-coupling pulse wings, explaining its weak pulse-area dependence. 
Consequently, the red Landau–Zener curves, based on the resonance crossing, need not coincide with the numerical contours obtained over the entire active window. Complete population inversion requires both sufficiently adiabatic evolution and adequate sweep coverage.

In addition, the sech and Gaussian envelopes exhibit similar overall chirp-dependent behavior, indicating that the sweep protocol plays a more important role than the envelope shape in determining the adiabatic operating region. The longer tails of the sech pulse extend the effective interaction window, whereas the Gaussian envelope confines the passage more strongly to the pulse center.
Thus, the envelope mainly affects the detailed dressed-state evolution, while the chirp largely determines the extent of the adiabatic operating region.

\section{Chirp-stabilized charging performance}
\label{Chirped-stabilized charging performance}
In this section, we investigate how the adiabatic passage discussed above affects the charging performance of the two-level battery by analyzing the stored energy, ergotropy, and charging power. We also compare the chirped protocols with a counterdiabatic driving protocol and examine their robustness against static detuning and dissipation.

\subsection{Figures of merit}
\label{sec.merit}

Population transfer from the state $|g\rangle$ to $|e\rangle$ changes both the energy stored in the two-level battery and the amount of work that can be extracted from its state.
The charging performance is characterized by the stored energy, ergotropy, ergotropy-to-energy ratio, and average charging power. Since the relevant energy observable is the bare Hamiltonian $H_b$, the energy increase relative to the initial state is defined as
\begin{equation}
	\Delta E(t)=\Tr[H_b\rho_s(t)]-\Tr[H_b\rho_s(t_{i})].
	\label{eq:stored}
\end{equation}
The second law of thermodynamics imposes a fundamental upper bound on the work that can be extracted from the stored energy. This maximum extractable work, called ergotropy, is defined as~\cite{PRE-042123-2013,RMP-031001-2024,Allahverdyan-2004}
\begin{equation}
	\begin{aligned}
		\varepsilon(t)&=\Tr[H_b\rho_s(t)]-
		\min_{U\in\mathcal{U}_{\rm cyc}}\Tr[H_bU\rho_s(t)U^\dagger]\\
		&=\Tr[H_b\rho_s(t)]-\Tr[H_b\tilde{\rho}_{s}(t)],
		\label{eq.ergotropy}
	\end{aligned}
\end{equation}
where $\mathcal{U}_{\rm cyc}$ denotes the set of cyclic unitary transformations, and $\tilde{\rho}_{s}(t)$ is the passive state associated with $\rho_s(t)$~\cite{Pusz-1978}.

To quantify the fraction of the stored energy that is extractable, we introduce the ergotropy-to-energy ratio~\cite{PRA-033715-2021}
\begin{equation}
	R_B(t)=\frac{\varepsilon(t)}{\Delta E(t)}.
	\label{eq.RB}
\end{equation}
Equation~(\ref{eq.RB}) is undefined at $\Delta E=0$, and hence we report this ratio only for
$\Delta E/(\hbar\omega_0)>10^{-6}$. 
In addition to these energetic figures of merit, the charging speed matters for practical operation. Therefore, we define the time-dependent average charging power
\begin{equation}
	P_B(t)=\frac{\Delta E(t)}{t-t_i}.
	\label{eq:average_power}
\end{equation}
For the initial condition $\rho_s(t_i)=|g\rangle\langle g|$, we have $\rho_{ee}(t_i)=0$. The corresponding stored energy can be written as
\begin{equation}
	\Delta E(t)=\hbar\omega_0\rho_{ee}(t),
	\label{eq:stored_two_level}
\end{equation}
and the ergotropy becomes
\begin{equation}
	\begin{aligned}
		\varepsilon(t)=\frac{\hbar\omega_{0}}{2}\left[X(t)+\sqrt{X^2(t)+4|\rho_{eg}(t)|^2}\right],
	\end{aligned}
	\label{eq:ergo_two_level}
\end{equation}
where $ X(t)=2\rho_{ee}(t)-1$.
Equation~\eqref{eq:ergo_two_level} follows directly from Eq.~\eqref{eq.ergotropy}, since the passive state of a two-level system is obtained by assigning the larger eigenvalue of $\rho_s(t)$ to the lower bare energy $-\hbar\omega_0/2$. The eigenvalues of $\rho_s(t)$ are $[1\pm\sqrt{X^2(t)+4|\rho_{eg}|^2}]/2$, which gives Eq.~\eqref{eq:ergo_two_level}. This expression shows explicitly that the coherence $|\rho_{eg}|$ can contribute to the extractable work.

In addition, to quantify how closely the actual open-system state follows the target instantaneous dressed state, we introduce the adiabatic-following fidelity
\begin{equation}
	F_{\rm ad}(t)=\langle\chi_{\rm target}(t)|\rho_s(t)|\chi_{\rm target}(t)\rangle,
	\label{eq:instantaneous_ad_fidelity}
\end{equation}
where the target branch is $|\chi_{\rm target}(t)\rangle=|\chi_-(t)\rangle$, since the system starts in $|g\rangle$ and the detuning changes from positive to negative values. Therefore, $F_{\rm ad}$ directly measures how closely the dissipative evolution follows the target dressed-state branch.

\subsection{Stored energy, ergotropy and adiabatic-following fidelity}
\label{sec:stored_energy_results}
To evaluate the charging performance, we exhibit the charging performance for the sech and Gaussian pulse envelopes in Figs.~\ref{fig:sech_triptych} and~\ref{fig:gaussian_triptych}. The unchirped reference curves exhibit the usual Rabi-oscillation behavior for both pulse envelopes. For an ideal resonant pulse with area $\stixcal{A}=\int\Omega(t)dt$, the excited-state population is given by~\cite{LJ-1975}
\begin{equation}
P_e=\sin^2\left(\frac{\stixcal{A}}{2}\right).
\label{eq:rabi_population}
\end{equation}
Equation~(\ref{eq:rabi_population}) shows that complete population inversion occurs at $\stixcal{A}=(2n+1)\pi$, while the population returns to the ground state at $\stixcal{A}=2n\pi$, with $n$ being a nonnegative integer. As $\stixcal{A}/\pi$ increases, $\Delta E$ and $\varepsilon$ oscillate rather than increase monotonically. Therefore, the unchirped protocol requires accurate pulse-area control. Deviations from the optimal pulse areas cause over-rotation, population back-transfer, and energy backflow.
\begin{figure*}[t]
	\centering
	\includegraphics[width=\textwidth]{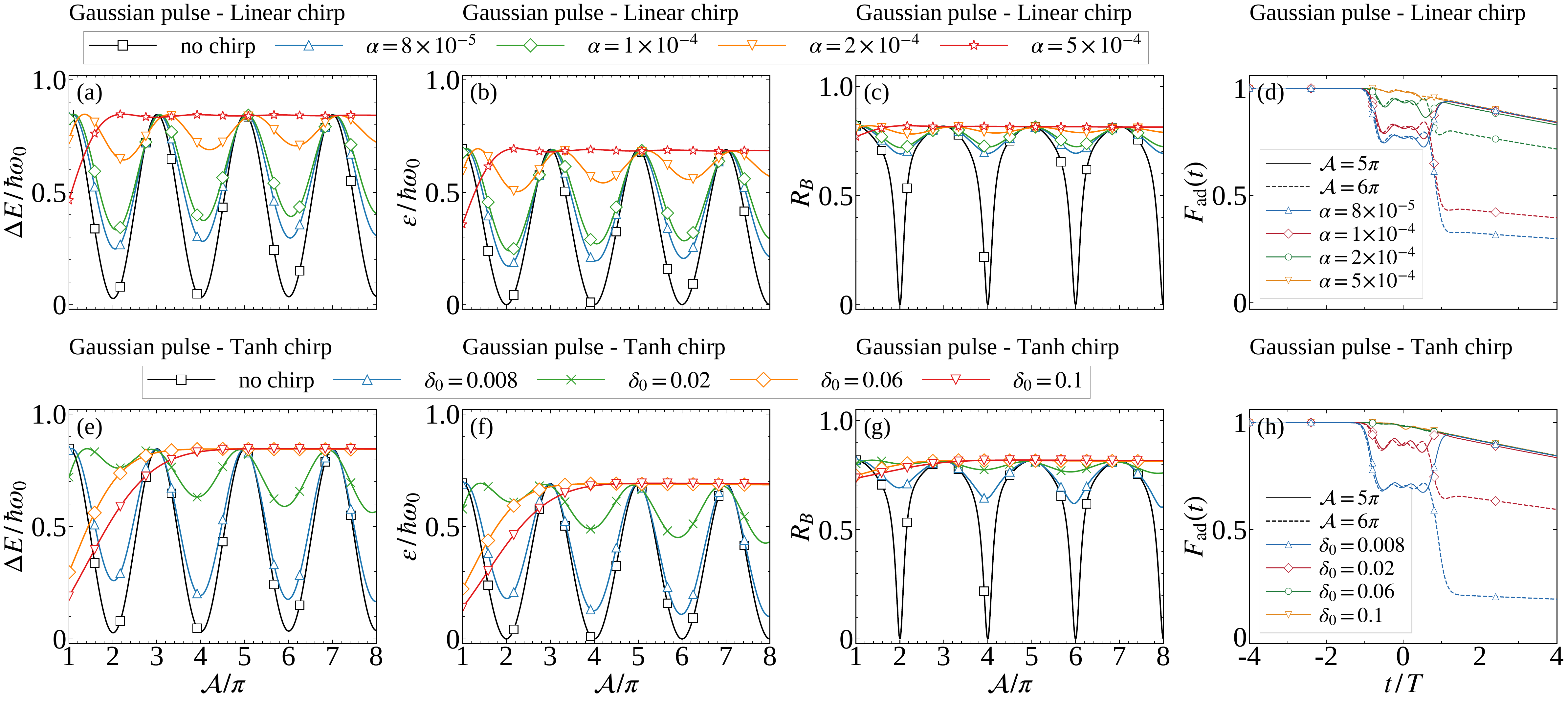}
	\caption{Charging performance of a Gaussian-driven two-level battery. (a--c) $\Delta E/\hbar\omega_0$, $\varepsilon/\hbar\omega_0$, and $R_{B}$ versus $\stixcal{A}/\pi$ for the linear chirp $\delta^{\rm (lin )}(t)$. (e)--(g) The same quantities for the tanh chirp $\delta^{\rm (tanh)}(t)$. The no-chirp result is plotted for comparison. (d,~h) Adiabatic-following fidelity $F_{\mathrm{ad}}(t)$ for selected chirp strengths at $\stixcal{A}=5\pi$~(solid) and
		$\stixcal{A}=6\pi$~(dashed). Other parameters are the same as those in Fig.~\ref{fig:sech_triptych}. }
	\label{fig:gaussian_triptych}
\end{figure*}

In contrast to the unchirped case, the chirp modulation fundamentally changes this behavior. For weak chirping, the minima of $\Delta E$ and $\varepsilon$ are lifted, but the Rabi-type oscillations remain visible, as seen for $\alpha=8\times10^{-5}$ and $1\times10^{-4}$ in Figs.~\ref{fig:sech_triptych}(a,~b) and~\ref{fig:gaussian_triptych}(a,~b) and for $\delta_0=0.008$ and $0.02$ in Figs.~\ref{fig:sech_triptych}(e,~f) and~\ref{fig:gaussian_triptych}(e,~f). In this weak-chirp regime, the pulse-area-controlled Rabi rotation remains significant. As the chirp strength increases further, the Rabi oscillations are progressively suppressed, and the stored energy and ergotropy become much less sensitive to the pulse area over a broad parameter range.
This behavior gives rise to broad charging plateaus, most clearly for $\alpha=5\times10^{-4}$ and for $\delta_0=0.06$ and $0.1$. In this regime, the frequency sweep drives the system through resonance, and the population transfer is governed by adiabatic following rather than by a precise rotation angle.

The same stabilization is also reflected in the ergotropy-to-energy ratio $R_{B}$. As shown in Figs.~\ref{fig:sech_triptych}(c, g) and \ref{fig:gaussian_triptych}(c, g), the strong pulse-area dependence of $R_{B}$ is substantially suppressed in the strong-chirp regime. The stored energy therefore remains largely extractable over the broad charging plateaus, which is important because a high stored energy alone does not necessarily imply a large amount of extractable work.

The adiabatic-following fidelity in Figs.~\ref{fig:sech_triptych}(d, h) and \ref{fig:gaussian_triptych}(d, h) further reveals the dynamical origin of the charging plateaus. For the linear chirp, the smaller rates $\alpha=8\times10^{-5}$ and $1\times10^{-4}$ show noticeable deviations from the target dressed-state branch, with the two pulse areas giving visibly different fidelity curves after the resonance passage. This pulse-area dependence is reduced at $\alpha=2\times10^{-4}$. At $\alpha=5\times10^{-4}$, the fidelity remains high for both $\stixcal{A}=5\pi$ and $6\pi$ over the main interaction region, and the corresponding curves show only a small separation. A similar trend is observed for the tanh chirp. For $\delta_0=0.008$, $F_{\rm ad}(t)$ remains strongly dependent on the pulse area, whereas increasing $\delta_0$ to $0.02$ only partially reduces this dependence. At $\delta_0=0.06$ and $0.1$, the fidelity remains high over the main interaction region for both $\stixcal{A}=5\pi$ and $6\pi$, and the corresponding curves show only a weak dependence on the pulse area for both pulse envelopes. At later times, $F_{\rm ad}(t)$ gradually decreases even in these strong-chirp cases, consistent with the dissipative evolution after the resonance passage.

These results indicate that the suppression of pulse-area sensitivity is governed mainly by the chirp strength, while the dependence on the pulse envelope remains comparatively weak. Therefore, for the two pulse envelopes considered here, robust charging is primarily controlled by the chirp rather than by the detailed envelope shape.

\subsection{Average charging power}
\label{sec:power_results}

We next examine how chirping affects the charging speed and the time-dependent average charging power. Figure~\ref{fig:average_power_chirped_protocols} shows the time-dependent average charging power $P_B(t)$ defined in Eq.~\eqref{eq:average_power} for pulse areas $\stixcal{A}=5\pi$ and $6\pi$. The pulse is centered at $t=0$, whereas the charging window begins at $t_i=-4T$. Thus, although the horizontal axis represents the centered time $t/T$, the averaging time entering $P_B(t)$ is $t-t_i=t+4T$.
\begin{figure*}[t]
	\centering
	\includegraphics[width=\textwidth]{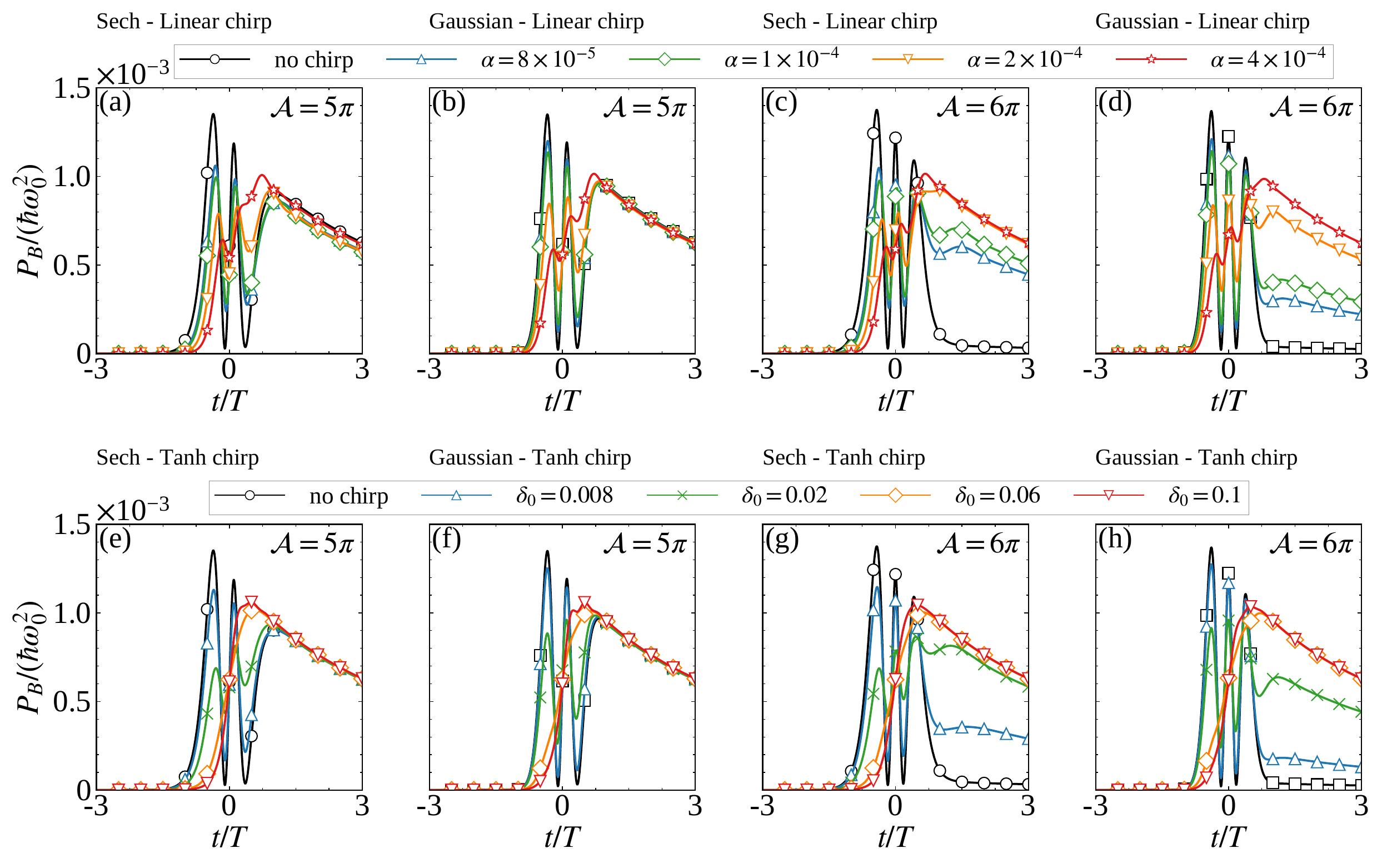}
	\caption{Normalized time-dependent average power $P_B/(\hbar\omega_0^2)$ versus the centered reduced time $t/T$. The upper and lower rows correspond to the linear and tanh chirps, respectively. Linear chirps are used in (a), (b), (c), (d), whereas tanh chirps are used in (e), (f), (g) and (h). Pulse areas are $\stixcal{A}=5\pi$ in (a), (b), (e), and (f), and $\stixcal{A}=6\pi$ in (c), (d), (g), and (h). The black curves denote the unchirped pulse, whereas the colored curves represent different chirp strengths, characterized by $\alpha$ for the linear chirp and by $\delta_0$ for the tanh chirp. Other parameters are the same as those in Fig.~\ref{fig:sech_triptych}.}
	\label{fig:average_power_chirped_protocols}
\end{figure*}

For the unchirped pulse, the power dynamics show a strong dependence on the pulse area. At $\stixcal{A}=5\pi$, resonant Rabi rotation produces a sharp maximum of $P_B(t)$ around the main interaction region, followed by a finite late-time value as the population remains largely in the excited state. At $\stixcal{A}=6\pi$, a similar transient maximum is followed by a rapid decrease of $P_B(t)$, as the subsequent Rabi rotation transfers the population back toward the ground state [Figs.~\ref{fig:average_power_chirped_protocols}(c,~d,~g,~h)]. This behavior is consistent with the strong pulse-area dependence of the stored energy and ergotropy discussed above.

Chirping gradually changes this behavior. For the linear sweep, the smaller chirp rates $\alpha=8\times10^{-5}$ and $1\times10^{-4}$ retain relatively sharp power structures around the pulse center, whereas the response becomes broader for $\alpha=2\times10^{-4}$ and $4\times10^{-4}$. A similar trend appears for the tanh sweep: the weak-modulation cases $\delta_0=0.008$ and $0.02$ retain pronounced Rabi-like features, while $\delta_0=0.06$ and $0.1$ produce broader power profiles extending over a longer time interval. The difference is particularly evident at $\stixcal{A}=6\pi$, where sufficiently strong chirping suppresses the rapid post-pulse decrease found in the unchirped case and substantially reduces the dependence of $P_B(t)$ on the pulse area.

The sech and Gaussian envelopes exhibit the same overall chirp-dependent behavior. The longer tails of the sech pulse lead to a somewhat broader temporal response, whereas the Gaussian pulse confines the main charging dynamics more closely to the pulse center. These differences become less pronounced in the strong-chirp regime, further indicating that the chirp strength plays the dominant role in suppressing pulse-area sensitivity. The resulting power dynamics shows that robust chirped charging may reduce the transient peak average power while providing a broader response and suppressing population back-transfer.

\subsection{Comparison of chirped-pulse charging with counterdiabatic driving and robustness against detuning and dissipation}
\label{sec:benchmark_detuning_dissipation}

The preceding analysis compares four chirp-only protocols with an unchirped resonant reference. To further assess the performance of the chirped protocols, we compare them in Fig.~\ref{fig:benchmark_detuning_dissipation} with an STA reference based on transitionless counterdiabatic driving~\cite{Berry-2009,Chen-2010}. For the two-level Hamiltonian in Eq.~\eqref{eq:heff}, the counterdiabatic term that cancels the nonadiabatic coupling in Eq.~\eqref{eq:Had} reads
\begin{equation}
H_{\rm CD}(t)=\frac{\hbar}{2}\dot{\theta}(t)\sigma_y .
\label{eq:counterdiabatic_hamiltonian}
\end{equation}
This counterdiabatic protocol requires an additional drive quadrature and hence is interpreted as a reference rather than as a resource-equivalent chirp-only protocol. Figure~\ref{fig:benchmark_detuning_dissipation}(a) compares the final ergotropy as a function of the pulse area at $\Delta=0$ for the unchirped sech pulse, the sech+tanh and sech+linear chirped pulses, and the CD/STA reference.

As expected for transitionless following, the CD/STA protocol strongly suppresses the pulse-area oscillations. The sech+tanh chirped protocol approaches this behavior over the main plateau region while requiring only a frequency sweep rather than an added $y$-quadrature counterdiabatic field. The sech+linear protocol also suppresses the pulse-area oscillations and yields a broad ergotropy plateau, while the sech+tanh protocol remains the closer chirp-only analogue of the CD/STA benchmark.
In contrast, the unchirped pulse remains governed by the Rabi oscillations, and the final ergotropy alternates between high and low values as the pulse area changes.
\begin{figure*}[t]
	\centering
	\includegraphics[width=1.0\textwidth]{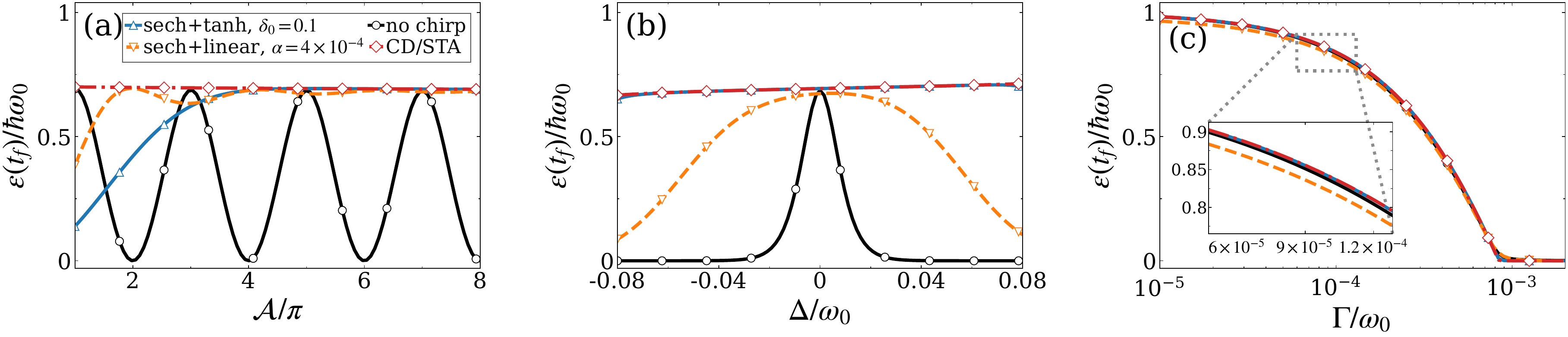}
	\caption{Benchmark robustness for selected charging protocols. (a) Final scaled ergotropy~$\varepsilon(t_{f})/(\hbar \omega_{0})$ as a function of $\mathcal{A}/\pi$ at $\Delta=0$ for an unchirped sech pulse, the sech+tanh and sech+linear chirped pulses, with the counterdiabatic shortcut-to-adiabaticity (CD/STA) benchmark included for reference. (b) Final $\varepsilon(t_{f})/(\hbar \omega_{0})$ versus the detuning $\Delta$ at fixed $\mathcal{A}=5\pi$. (c) Final $\varepsilon(t_{f})/(\hbar \omega_{0})$ versus the common dissipative scale $\Gamma_\downarrow/\omega_0=\gamma_\phi/\omega_0$. Other parameters are $T=200\omega_0^{-1}$ and $\beta=5(\hbar\omega_0)^{-1}$.
		For panels (a) and (b), $\Gamma_{\downarrow}=\gamma_{\phi}=2\times10^{-4}\omega_0$ and $\Gamma_{\uparrow}=\Gamma_{\downarrow}e^{-5}$.
		In panel (c), the common dissipative rate $\Gamma_{\downarrow}=\gamma_{\phi}\equiv\Gamma$ is varied, with $\Gamma_{\uparrow}=\Gamma_{\downarrow}e^{-5}$ maintained throughout.}
	\label{fig:benchmark_detuning_dissipation}
\end{figure*}

To examine robustness against static detuning errors, Figure~\ref{fig:benchmark_detuning_dissipation}(b) shows the final ergotropy versus static detuning $\Delta$ at fixed $\stixcal {A}=5\pi$. For a chirped pulse,  robust population transfer is maintained as long as the shifted crossing remains within the active interaction window and the local gap remains sufficiently large. The tanh-chirped protocol retains high ergotropy over a finite interval of $\Delta$, thereby exhibiting the robustness expected from adiabatic rapid passage. By contrast, the unchirped reference has no sweep-coverage mechanism and is more sensitive to deviations from exact resonance.

Figure~\ref{fig:benchmark_detuning_dissipation}(c) further shows that the final ergotropy decreases as the dissipative rates increase. This degradation arises from dephasing during the passage and relaxation of the excited-state population. The weak-error conditions discussed in Sec.~\ref{app:validity} characterize the asymptotic regime of nearly ideal charging and are not required to hold throughout the plotted parameter range.
As the dissipative rates increase, the final ergotropy decreases because the dephasing reduces the coherence generated near the avoided crossing, while the relaxation removes excited-state population after the passage. A similar dependence on dissipation is found for the Gaussian pulse~(not shown).

\section{Discussions}
\label{Discussion}
In this section, we discuss the experimental feasibility of the chirped-pulse protocol and establish the pulse-parameter design ranges for near-ideal adiabatic charging.
\subsection{Experimental feasibility}
\label{sec:platform_mapping}

A promising experimental platform for implementing the proposed charging protocol is superconducting circuits~\cite{QST-045018-2022}. Such systems enable precise control of the microwave amplitude, phase, duration, and frequency. In particular, quantum-battery charging has already been demonstrated using a superconducting transmon qutrit~\cite{QST-045018-2022}. We adopt a representative two-level transition frequency $\omega_0=2\pi\times5.0~{\rm GHz}$, typical for transmons and circuit-QED devices~\cite{Koch-2007,Kjaergaard-2020,Blais-2021,Krantz-2019}. For this transition frequency, the characteristic time scale is
$\omega_0^{-1}=31.8~{\rm ps}$. Accordingly, $T=200\omega_0^{-1}$ corresponds to $6.37~{\rm ns}$, and the full charging window $[-4T,4T]$ spans $50.9~{\rm ns}$. The dimensionless thermal parameter $\beta\hbar\omega_0=5$ yields $T_{\rm bath}=\hbar\omega_0/(5k_B)\simeq48~{\rm mK}$. Table~\ref{tab:sc_mapping} summarizes these parameter conversions.

The pulse durations and chirp bandwidths considered here are experimentally accessible in superconducting circuits. Nevertheless, the largest peak Rabi frequencies $\Omega_{\rm max}$ impose stringent requirements on weakly anharmonic transmons and may require optimized pulse shaping to suppress leakage to higher energy levels~\cite{Motzoi-2009}. The required $\Omega_{\rm max}$ can be reduced by increasing the pulse duration $T$ while keeping the pulse area fixed, with chirp parameters adjusted accordingly to maintain the adiabatic-passage conditions. The dissipative rates adopted in our simulations correspond to relaxation and dephasing times on the sub-microsecond scale. State-of-the-art superconducting qubits routinely exhibit much longer coherence times, ranging from tens of microseconds to above $0.3~{\rm ms}$ for high-coherence materials~\cite{Kjaergaard-2020,Place-2021}. Therefore, the dimensionless ratios $\Gamma_\downarrow/\omega_0$ and $\gamma_\phi/\omega_0$
can be smaller than our simulated values in low-loss devices, whereas stronger effective dissipation can be engineered via Purcell enhancement or coupling to a lossy auxiliary mode.
\begin{widetext}
	\begingroup
	\small
	\refstepcounter{table}\label{tab:sc_mapping}
	\noindent\textbf{TABLE~\thetable.} Parameters used in numerical simulations and their corresponding values for a superconducting-qubit implementation with
$\omega_0/2\pi=5.0~{\rm GHz}$. The parameter $n=\stixcal{A}/\pi$ specifies the pulse area in units of $\pi$.
	\begin{center}
		\begin{ruledtabular}
			\begin{tabular}{llll}
			\toprule
			{Notation}
			& {Remarks}
			& {Scaled parameters}
			& {Parameters} \\
			\midrule
			\hline

			$\omega_{0}$
			& transition frequency
			& $1$
			& $2\pi \times 5.0~\mathrm{GHz}$ \\

			$T$
			& pulse width
			& $T\omega_{0}=200$
			& $6.37~\mathrm{ns}$ \\

			$\Omega_{\max}$
			& peak Rabi frequency
			& $\Omega_{\max}/\omega_{0}=n\pi/200$, $n\in[1,8]$
			& $2\pi \times (78.5-628)~\mathrm{MHz}$ \\

			$\delta_{0}$
			& tanh-chirp amplitude
			& $\delta_{0}/\omega_{0}=0.005-0.14$
			& $2\pi \times (25-700)~\mathrm{MHz}$ \\

			$\alpha$
			& linear chirp rate
			& $\alpha/\omega_{0}^{2}
			=3\times10^{-5}-6\times10^{-4}$
			& $2\pi \times (4.7-94)~\mathrm{MHz/ns}$ \\

			$\Gamma_{\downarrow}$
			& relaxation rate
			& $\Gamma_{\downarrow}/\omega_{0}=2\times10^{-4}$
			& $2\pi \times 1.0~\mathrm{MHz}$ \\

			$\gamma_{\phi}$
			& pure-dephasing rate
			& $\gamma_{\phi}/\omega_{0}=2\times10^{-4}$
			& $2\pi \times 1.0~\mathrm{MHz}$ \\

			$\Gamma_{\uparrow}$
			& thermal-excitation rate
			& $\Gamma_{\uparrow}
			=\Gamma_{\downarrow}e^{-5}$
			& $2\pi \times 6.7~\mathrm{kHz}$ \\

			$T_{\mathrm{bath}}$
			& bath temperature
			& $\beta\hbar\omega_{0}=5$
			& $48~\mathrm{mK}$ \\
			\bottomrule
			\end{tabular}
		\end{ruledtabular}
	\end{center}
	\endgroup
\end{widetext}
\subsection{Adiabatic passage conditions and dissipative constraints}
\label{app:validity}

Beyond local adiabaticity, robust population transfer requires sufficient sweep coverage. The detuning must cross resonance and attain sufficiently large magnitudes at both boundaries of the effective interaction window, so that the target dressed state asymptotically approaches the bare ground and excited states, respectively. A practical criterion reads $\delta(t_i^{\rm eff})\delta(t_f^{\rm eff})<0$ and $|\delta(t_{i,f}^{\rm eff})|/\Omega(t_{i,f}^{\rm eff})\gtrsim q$, with $q\sim3$--$5$, where $t_{i,f}^{\rm eff}$ denote the boundaries of $I_{\rm active}$ defined in Eq.~(\ref{eq:active_window}). 

{Furthermore, to approach full charging through adiabatic evolution, both pure dephasing and relaxation impose additional constraints on the duration of the adiabatic passage. Near resonance, strong mixing between the bare states renders the target dressed state especially vulnerable to pure dephasing. Approximating $\Omega(t)\simeq\Omega_*$ and $\delta(t)\simeq-v_*(t-t_*)$, the effective duration of this strongly mixed region is estimated as
\begin{equation}
\
\tau_{\rm mix}\sim\int_{-\infty}^{\infty}\frac{\Omega_*^2}{\Omega_*^2+v_*^2t^2}\,dt
=\frac{\pi\Omega_*}{v_*}.
\
\end{equation}
Here, the integration variable is measured relative to the resonance crossing instant, and the local approximation requires $\Omega_*/v_*\ll T$. The associated pure-dephasing error is $p_\phi\sim(\gamma_\phi/2)\tau_{\rm mix}$. Relaxation errors accumulate according to the excited-state population along the target adiabatic branch, yielding $\tau_e=\int_{t_i}^{t_f}P_e^{\rm ad}(t)\,dt$, where $P_e^{\rm ad}(t)=[1-\delta(t)/\Upsilon(t)]/2$. For weak relaxation, $p_{\rm rel}\simeq\Gamma_\downarrow\tau_e$. Within the symmetric window $[-4T,4T]$ in $\Delta=0$, one obtains $\tau_e=4T$.

In the weak-error regime with negligible thermal excitation, the leading-order error budget reads $\eta_{\rm tot}\simeq p_{\rm na}+(\gamma_\phi/2)\tau_{\rm mix}+\Gamma_\downarrow\tau_e$, where $p_{\rm na}$ is given by Eq.~(\ref{eq:lz_probability}). Combined with adequate sweep coverage, the practical conditions for robust adiabatic charging are
\begin{subequations}
	\label{eq:plateau_conditions}
	\begin{align}
		\eta_{\rm tot}&\ll1,
		&\Lambda_{\max}^{\rm active}&\lesssim0.1,\\
		\frac{\Gamma_2}{\Upsilon_{\min}^{\rm active}}&\ll1,
		&\Gamma_\downarrow\tau_e&\ll1.
	\end{align}
\end{subequations}
 These conditions encapsulate a trade-off: slowing down the resonance crossing suppresses nonadiabatic transitions, whereas shortening the charging window mitigates dissipative errors. Thus, using Eq.~(\ref{eq:plateau_conditions}), we can establish an approximate design criterion for near-ideal adiabatic charging.

\section{Conclusion}
\label{conclusion}

We have studied the charging of a two-level quantum battery driven by four families of chirped pulses. Our results demonstrate that chirp modulation can switch the underlying charging mechanism from pulse-area-controlled Rabi rotation to frequency-swept adiabatic passage. Consequently, population back-transfer and energy backflow are suppressed, giving rise to broad plateaus of stored energy and ergotropy over an extended range of pulse areas. We have also established the conditions for robust adiabatic charging by accounting for sweep coverage, nonadiabatic transitions, and dissipative processes. Among the four protocols considered, the sech envelope combined with a tanh chirp achieves the best balance between plateau width, final ergotropy, and stability of charging speed.

We have also compared the chirped-pulse protocol, which requires no extra quadrature control field, with a counterdiabatic shortcut-to-adiabaticity reference. Moreover, chirped charging retains robustness against static detuning and weak-to-moderate dissipation. These findings establish chirped-pulse control as a viable route toward robust quantum-battery charging in dissipative environments.

\begin{acknowledgments}
J.-Q.L. was supported in part by the National Natural Science Foundation of China (Grants No.~12175061, No.~12575015, No.~12247105, No.~11935006, and No.~12421005), the National Key Research and Development Program of China (Grant No.~2024YFE0102400), and the Hunan Provincial Major Sci-Tech Program (Grant No.~2023ZJ1010). J.-F. H. is partially supported by the National Natural Science Foundation of China (Grant No. 12475016) and the Science and Technology Innovation Program of Hunan Province (Grant No. 2025RC3141).
X.-L. Yin is partially supported by the National Natural Science Foundation of China (Grant No.~12605031).
\end{acknowledgments}


\begin{thebibliography}{99}

\bibitem{PRE-042123-2013}
		R. Alicki and M. Fannes, Entanglement boost for extractable work from ensembles of quantum batteries, Phys. Rev. E \textbf{87}, 042123 (2013).

\bibitem{Hovhannisyan-2013}
K. V. Hovhannisyan, M. Perarnau-Llobet, M. Huber, and A. Ac\'in, Entanglement generation is not necessary for optimal work extraction, Phys. Rev. Lett. \textbf{111}, 240401 (2013).

\bibitem{CampaioliBook-2018}
F. Campaioli, F. A. Pollock, and S. Vinjanampathy, Quantum batteries, in \textit{Thermodynamics in the Quantum Regime: Fundamental Aspects and New Directions}, edited by F. Binder, L. A. Correa, C. Gogolin, J. Anders, and G. Adesso (Springer, Cham, 2018), pp. 207--225.

\bibitem{RMP-031001-2024}
		F. Campaioli, S. Gherardini, J. Q. Quach, M. Polini, and G. M. Andolina, Colloquium: Quantum batteries, Rev. Mod. Phys. \textbf{96}, 031001 (2024).

\bibitem{NJP-075015-2015}
		F. C. Binder, S. Vinjanampathy, K. Modi, and J. Goold, Quantacell: Powerful charging of quantum batteries, New J. Phys. \textbf{17}, 075015 (2015).

\bibitem{PRB-205423-2018}
		G. M. Andolina, D. Farina, A. Mari, V. Pellegrini, V. Giovannetti, and M. Polini, Charger-mediated energy transfer in exactly solvable models for quantum batteries, Phys. Rev. B \textbf{98}, 205423 (2018).

\bibitem{PRL-047702-2019}
G. M. Andolina, M. Keck, A. Mari, M. Campisi, V. Giovannetti, and M. Polini, Extractable work, the role of correlations, and asymptotic freedom in quantum batteries, Phys. Rev. Lett. \textbf{122}, 047702 (2019).

\bibitem{PRR-023113-2020}
S. Juli\`a-Farr\'e, T. Salamon, A. Riera, M. N. Bera, and M. Lewenstein, Bounds on the capacity and power of quantum batteries, Phys. Rev. Research \textbf{2}, 023113 (2020).

\bibitem{PRL-040601-2020}
L. P. Garc\'ia-Pintos, A. Hamma, and A. del Campo, Fluctuations in extractable work bound the charging power of quantum batteries, Phys. Rev. Lett. \textbf{125}, 040601 (2020).

\bibitem{PRL-130602-2022}
H.-L. Shi, S. Ding, Q.-K. Wan, X.-H. Wang, and W.-L. Yang, Entanglement, coherence, and extractable work in quantum batteries, Phys. Rev. Lett. \textbf{129}, 130602 (2022).

\bibitem{Wen-2025-wireless}
J. Wen, Z. Wen, P. Peng, and G.-Q. Li, Continuous-variable quantum battery with wireless and remote charging, Front. Phys. \textbf{21}, 102201 (2026).

\bibitem{Beder-2026-cavities}
I. Beder, D. Ferraro, and P. A. Brand\~ao, Work extraction from a quantum battery charged through an array of coupled cavities, Phys. Rev. A \textbf{113}, 022210 (2026).

\bibitem{PRL-117702-2018}
		D. Ferraro, M. Campisi, G. M. Andolina, V. Pellegrini, and M. Polini, High-Power Collective Charging of a Solid-State Quantum Battery, Phys. Rev. Lett. \textbf{120}, 117702 (2018).

\bibitem{PRL-150601-2017}
		F. Campaioli, F. A. Pollock, F. C. Binder, L. Céleri, J. Goold, S. Vinjanampathy, and K. Modi, Enhancing the Charging Power
		of Quantum Batteries, Phys. Rev. Lett. \textbf{118}, 150601 (2017).

\bibitem{PRA-022106-2018}
		T. P. Le, J. Levinsen, K. Modi, M. M. Parish, and F. A. Pollock, Spin-chain model of a many-body quantum battery, Phys. Rev. A \textbf{97}, 022106 (2018).

\bibitem{PRB-205437-2019}
		G. M. Andolina, M. Keck, A. Mari, V. Giovannetti, and M. Polini, Quantum versus classical many-body batteries, Phys.
		Rev. B \textbf{99}, 205437 (2019).

\bibitem{Ghosh-2020-interactions}
S. Ghosh, T. Chanda, and A. Sen(De), Enhancement in the performance of a quantum battery by ordered and disordered interactions, Phys. Rev. A \textbf{101}, 032115 (2020).

\bibitem{Rossini-2019-MBL}
D. Rossini, G. M. Andolina, and M. Polini, Many-body localized quantum batteries, Phys. Rev. B \textbf{100}, 115142 (2019).

\bibitem{Jing-2021-PRA}
       Y. Huangfu and J. Jing, High-capacity and high-power collective charging with spin chargers, Phys. Rev. E 104, 024129 (2021).

\bibitem{NJP-083085-2020}
M. Carrega, A. Crescente, D. Ferraro, and M. Sassetti, Dissipative dynamics of an open quantum battery, New J. Phys. \textbf{22}, 083085 (2020).

\bibitem{PRA-033715-2021}
		F. Zhao, F.-Q. Dou, and Q. Zhao, Quantum battery of interacting spins with environmental noise, Phys. Rev. A \textbf{103}, 033715 (2021).

\bibitem{PRA-023725-2023}
F.-Q. Dou and F.-M. Yang, Superconducting transmon qubit-resonator quantum battery, Phys. Rev. A \textbf{107}, 023725 (2023).

\bibitem{PRA-032211-2024}
D. Zhang, S. Ma, Y. Jiang, Y. Yu, G. Jin, and A. Chen, Quantum battery with interactive atomic collective charging, Phys. Rev. A \textbf{110}, 032211 (2024).

\bibitem{PRL-130401-2025}
S. Pokhrel and J. Gea-Banacloche, Large collective power enhancement in dissipative charging of a quantum battery, Phys. Rev. Lett. \textbf{134}, 130401 (2025).

\bibitem{NJP-083007-2020}
F. H. Kamin, F. T. Tabesh, S. Salimi, F. Kheirandish, and A. C. Santos, Non-Markovian effects on charging and self-discharging process of quantum batteries, New J. Phys. \textbf{22}, 083007 (2020).

\bibitem{QST-035007-2023}
D. Morrone, M. A. C. Rossi, A. Smirne, and M. G. Genoni, Charging a quantum battery in a non-Markovian environment: A collisional model approach, Quantum Sci. Technol. \textbf{8}, 035007 (2023).

\bibitem{PRA-032202-2026}
J.-K. Xu, J.-B. You, and W.-L. Yang, Non-Markovian-assisted advantage for central-spin quantum battery, Phys. Rev. A \textbf{113}, 032202 (2026).

\bibitem{Pirmoradian-2019-aging}
F. Pirmoradian and K. M{\o}lmer, Aging of a quantum battery, Phys. Rev. A \textbf{100}, 043833 (2019).

\bibitem{Santos-2021-self-discharge}
A. C. Santos, Quantum advantage of two-level batteries in the self-discharging process, Phys. Rev. E \textbf{103}, 042118 (2021).

\bibitem{PRL-220402-2025}
I. Medina, O. Culhane, F. C. Binder, G. T. Landi, and J. Goold, Anomalous discharging of quantum batteries: The ergotropic Mpemba effect, Phys. Rev. Lett. \textbf{134}, 220402 (2025).

\bibitem{PRL-020405-2025}
W.-L. Song, J.-L. Wang, B. Zhou, W.-L. Yang, and J.-H. An, Self-discharging mitigated quantum battery, Phys. Rev. Lett. \textbf{135}, 020405 (2025).

\bibitem{PRL-210601-2019}
		F. Barra, Dissipative Charging of a Quantum Battery, Phys. Rev. Lett. \textbf{122}, 210601 (2019).

\bibitem{PRApplied-024092-2020}
J. Q. Quach and W. J. Munro, Using dark states to charge and stabilize open quantum batteries, Phys. Rev. Applied \textbf{14}, 024092 (2020).

\bibitem{PRL-210402-2024}
B. Ahmadi, P. Mazurek, P. Horodecki, and S. Barzanjeh, Nonreciprocal quantum batteries, Phys. Rev. Lett. \textbf{132}, 210402 (2024).

\bibitem{PRA-012617-2026}
J. Dias, H. Wang, K. Nemoto, F. Nori, and W. J. Munro, Efficient charging of multiple open quantum batteries through dissipation and pumping, Phys. Rev. A \textbf{113}, 012617 (2026).

\bibitem{Guo-2026-waveguide}
M. Guo, J. Huang, R.-Y. Gong, X.-L. Yin, and G. Zhang, Phase-tunable remote nonreciprocal charging in waveguide QED, arXiv:2605.21909.

\bibitem{PRA-060201-2020}
S.-Y. Bai and J.-H. An, Floquet engineering to reactivate a dissipative quantum battery, Phys. Rev. A \textbf{102}, 060201(R) (2020).

\bibitem{PRL-090401-2024}
W.-L. Song, H.-B. Liu, B. Zhou, W.-L. Yang, and J.-H. An, Remote charging and degradation suppression for the quantum battery, Phys. Rev. Lett. \textbf{132}, 090401 (2024).

\bibitem{Lu-2025-topological}
Z.-G. Lu, G. Tian, X.-Y. L\"u, and C. Shang, Topological quantum batteries, Phys. Rev. Lett. \textbf{134}, 180401 (2025).

\bibitem{Yan-2026-giant-atom}
K.-X. Yan, Y. Liu, Y. Xiao, J.-H. Lin, J. Song, Y.-H. Chen, F. Nori, and Y. Xia, Giant-atom quantum batteries: Lossless energy transfer via interference engineering, Phys. Rev. Lett. \textbf{136}, 180401 (2026).

\bibitem{PRR-013095-2020}
S. Gherardini, F. Campaioli, F. Caruso, and F. C. Binder, Stabilizing open quantum batteries by sequential measurements, Phys. Rev. Research \textbf{2}, 013095 (2020).

\bibitem{Quantum-500-2021}
M. T. Mitchison, J. Goold, and J. Prior, Charging a quantum battery with linear feedback control, Quantum \textbf{5}, 500 (2021).

\bibitem{PRE-014138-2022}
Y. Yao and X. Q. Shao, Optimal charging of open spin-chain quantum batteries via homodyne-based feedback control, Phys. Rev. E \textbf{106}, 014138 (2022).

\bibitem{PRR-L032035-2026}
X.-L. Yin, M. Guo, J. Huang, H.-w. J. Lee, and G. Zhang, Feedback-enhanced driven-dissipative quantum batteries in waveguide QED systems, Phys. Rev. Research \textbf{8}, L032035 (2026).

\bibitem{PRL-050406-2026}
G. Cenedese, G. Benenti, D. Ferraro, and M. G. Genoni, Boosting work extraction in quantum batteries via continuous environment monitoring, Phys. Rev. Lett. \textbf{137}, 050406 (2026).

\bibitem{SCA-2022}
		J. Q. Quach, K. E. McGhee, L. Ganzer, D. M. Rouse, B. W. Lovett, E. M. Gauger, J. Keeling, G. Cerullo, D. G. Lidzey, and T. Virgili, Superabsorption in an organic microcavity: Toward a quantum battery, Sci. Adv. \textbf{8}, eabk3160 (2022).

\bibitem{QST-045018-2022}
		C.-K. Hu, J. Qiu, P. J. P. Souza, J. Yuan, Y. Zhou, L. Zhang, J. Chu, X. Pan, L. Hu, J. Li, Y. Xu, Y. Zhong, S. Liu, F. Yan, D. Tan, R. Bachelard, C. J. Villas-Boas, A. C. Santos, and D. Yu, Optimal charging of a superconducting quantum battery, Quantum Sci. Technol. \textbf{7}, 045018 (2022).

\bibitem{PRR-023091-2024}
G. Gemme, M. Grossi, S. Vallecorsa, M. Sassetti, and D. Ferraro, Qutrit quantum battery: Comparing different charging protocols, Phys. Rev. Research \textbf{6}, 023091 (2024).

\bibitem{PRL-060401-2026}
C.-K. Hu, C. Liu, J. Zhao, L. Zhong, Y. Zhou, M. Liu, H. Yuan, Y. Lin, Y. Xu, G. Hu, G. Xie, Z. Liu, R. Zhou, Y. Ri, W. Zhang, R. Deng, A. Saguia, X. Linpeng, M. S. Sarandy, S. Liu, A. C. Santos, D. Tan, and D. Yu, Quantum charging advantage in superconducting solid-state batteries, Phys. Rev. Lett. \textbf{136}, 060401 (2026).

\bibitem{PRA-042601-2022}
J. Joshi and T. S. Mahesh, Experimental investigation of a quantum battery using star-topology NMR spin systems, Phys. Rev. A \textbf{106}, 042601 (2022).

\bibitem{PRA-L03201-2023}
		X. Huang, K. Wang, L. Xiao, L. Gao, H. Lin, and P. Xue, Demonstration of the charging progress of quantum batteries, Phys. Rev. A \textbf{107}, L030201 (2023).

\bibitem{PRE-052106-2019}
		Y.-Y. Zhang, T.-R. Yang, L. Fu, and X. Wang, Powerful harmonic charging in a quantum battery, Phys. Rev. E \textbf{99}, 052106 (2019).

\bibitem{NJP-063057-2020}
		A. Crescente, M. Carrega, M. Sassetti, and D. Ferraro, Charging and energy fluctuations of a driven quantum battery, New J. Phys. \textbf{22}, 063057 (2020).
        
\bibitem{Evangelakos-2024-optimal}
        V. Evangelakos, E. Paspalakis, and D. Stefanatos, Fast charging of an Ising-spin-pair quantum battery using optimal control, Phys. Rev. A \textbf{110}, 052601 (2024).

\bibitem{PRA-032218-2023}
F. Mazzoncini, V. Cavina, G. M. Andolina, P. A. Erdman, and V. Giovannetti, Optimal control methods for quantum batteries, Phys. Rev. A \textbf{107}, 032218 (2023).

\bibitem{APL-154002-2023}
		Y. Ge, X. Yu, W. Xin, Z. Wang, Y. Zhang, W. Zheng, S. Li, D. Lan, and Y. Yu, Efficient charging and discharging of a superconducting quantum battery through frequency-modulated driving, Appl. Phys. Lett. \textbf{123}, 154002 (2023).

\bibitem{EPL-10001-2024}
C. A. Downing and M. S. Ukhtary, Energetics of a pulsed quantum battery, Europhys. Lett. \textbf{146}, 10001 (2024).

\bibitem{FrontPhys-31503-2022}
F.-Q. Dou, Y.-J. Wang, and J.-A. Sun, Highly efficient charging and discharging of three-level quantum batteries through shortcuts to adiabaticity, Front. Phys. \textbf{17}, 31503 (2022).

\bibitem{PLA-130124-2025}
S.-F. Qi and J. Jing, Quantum recharging by shortcut to adiabaticity, Phys. Lett. A \textbf{530}, 130124 (2025).

\bibitem{PRB-035421-2019}
D. Farina, G. M. Andolina, A. Mari, M. Polini, and V. Giovannetti, Charger-mediated energy transfer for quantum batteries: An open-system approach, Phys. Rev. B \textbf{99}, 035421 (2019).

\bibitem{LJ-1975}
L. Allen and J. H. Eberly, \textit{Optical Resonance and Two-Level Atoms} (Wiley, New York, 1975).

\bibitem{ARP-2001}
		N. V. Vitanov, T. Halfmann, B. W. Shore, and K. Bergmann, Laser-induced population transfer by adiabatic passage techniques, Annu. Rev. Phys. Chem. \textbf{52}, 763 (2001).

\bibitem{PRA-023409-2002}
		S. Guérin, S. Thomas, and H. R. Jauslin, Optimization of population transfer by adiabatic passage, Phys. Rev. A \textbf{65}, 023409 (2002).

\bibitem{PRA-063422-2007}
		T. Lu, X. Miao, and H. Metcalf, Nonadiabatic transitions in finite-time adiabatic rapid passage, Phys. Rev. A \textbf{75}, 063422 (2007).

\bibitem{PRA-033411-2011}
		T. Lu, Population inversion by chirped pulses, Phys. Rev. A \textbf{84}, 033411 (2011).

\bibitem{PRE-032107-2019}
A. C. Santos, B. \c{C}akmak, S. Campbell, and N. T. Zinner, Stable adiabatic quantum batteries, Phys. Rev. E \textbf{100}, 032107 (2019).

\bibitem{GKSL-1976}
		{V. Gorini, A. Kossakowski, and E. C. G. Sudarshan, Completely positive dynamical semigroups of $N$-level systems, J. Math. Phys. \textbf{17}, 821 (1976).}

\bibitem{Lindblad-1976}
		{G. Lindblad, On the generators of quantum dynamical semigroups, Commun. Math. Phys. \textbf{48}, 119 (1976).}

\bibitem{BreuerPetruccione-2002}
		{H.-P. Breuer and F. Petruccione, \textit{The Theory of Open Quantum Systems} (Oxford University Press, Oxford, 2002).}

\bibitem{Hioe-1984}
		F. T. Hioe, Solution of Bloch equations involving amplitude and frequency modulations, Phys. Rev. A \textbf{30}, 2100 (1984).

\bibitem{Suominen-1992}
		K.-A. Suominen and B. M. Garraway, Population transfer in a level-crossing model with two time scales, Phys. Rev. A \textbf{45}, 374 (1992).

\bibitem{Torosov-2011}
		B. T. Torosov and N. V. Vitanov, Evolution of superpositions of quantum states through a level crossing, Phys. Rev. A \textbf{84}, 063411 (2011).

\bibitem{Davies-1974}
		{E. B. Davies, Markovian master equations, Commun. Math. Phys. \textbf{39}, 91 (1974).}

\bibitem{LevyKosloff-2014}
		{A. Levy and R. Kosloff, The local approach to quantum transport may violate the second law of thermodynamics, Europhys. Lett. \textbf{107}, 20004 (2014).}

\bibitem{Hofer-2018}
		{P. P. Hofer, M. Perarnau-Llobet, L. D. M. Miranda, G. Haack, R. Silva, J. B. Brask, and N. Brunner, Markovian master equations for quantum thermal machines: local versus global approach, New J. Phys. \textbf{19}, 123037 (2017).}

\bibitem{Dann-2018}
		{R. Dann, A. Levy, and R. Kosloff, Time-dependent Markovian quantum master equation, Phys. Rev. A \textbf{98}, 052129 (2018).}
\bibitem{Adiabatensatzes}		
         {M. Born and V. Fock, Beweis des Adiabatensatzes, Z.Phys. 51, 165 (1928).}
         
 \bibitem{TKato}        
         T. Kato, On the adiabatic theorem of quantum mechanics, J. Phys. Soc. Jpn. 5, 435 (1950).	
	
\bibitem{Landau-1932}
		{L. D. Landau, Zur Theorie der Energieuebertragung. II, Phys. Z. Sowjetunion \textbf{2}, 46 (1932).}

\bibitem{Zener-1932}
		{C. Zener, Non-adiabatic crossing of energy levels, Proc. R. Soc. Lond. A \textbf{137}, 696 (1932).}

\bibitem{Allahverdyan-2004}
		{A. E. Allahverdyan, R. Balian, and Th. M. Nieuwenhuizen, Maximal work extraction from finite quantum systems, Europhys. Lett. \textbf{67}, 565 (2004).}

\bibitem{Pusz-1978}
		{W. Pusz and S. L. Woronowicz, Passive states and KMS states for general quantum systems, Commun. Math. Phys. \textbf{58}, 273 (1978).}

\bibitem{Berry-2009}
		M. V. Berry, Transitionless quantum driving, J. Phys. A: Math. Theor. \textbf{42}, 365303 (2009).

\bibitem{Chen-2010}
		X. Chen, I. Lizuain, A. Ruschhaupt, D. Gu\'ery-Odelin, and J. G. Muga, Shortcut to adiabatic passage in two- and three-level atoms, Phys. Rev. Lett. \textbf{105}, 123003 (2010).

\bibitem{Koch-2007}
		{J. Koch, T. M. Yu, J. Gambetta, A. A. Houck, D. I. Schuster, J. Majer, A. Blais, M. H. Devoret, S. M. Girvin, and R. J. Schoelkopf, Charge-insensitive qubit design derived from the Cooper pair box, Phys. Rev. A \textbf{76}, 042319 (2007).}

\bibitem{Kjaergaard-2020}
		{M. Kjaergaard, M. E. Schwartz, J. Braum{\"u}ller, P. Krantz, J. I.-J. Wang, S. Gustavsson, and W. D. Oliver, Superconducting qubits: Current state of play, Annu. Rev. Condens. Matter Phys. \textbf{11}, 369 (2020).}

\bibitem{Blais-2021}
		{A. Blais, A. L. Grimsmo, S. M. Girvin, and A. Wallraff, Circuit quantum electrodynamics, Rev. Mod. Phys. \textbf{93}, 025005 (2021).}

\bibitem{Krantz-2019}
		{P. Krantz, M. Kjaergaard, F. Yan, T. P. Orlando, S. Gustavsson, and W. D. Oliver, A quantum engineer's guide to superconducting qubits, Appl. Phys. Rev. \textbf{6}, 021318 (2019).}

\bibitem{Motzoi-2009}
		F. Motzoi, J. M. Gambetta, P. Rebentrost, and F. K. Wilhelm, Simple pulses for elimination of leakage in weakly nonlinear qubits, Phys. Rev. Lett. \textbf{103}, 110501 (2009).

\bibitem{Place-2021}
		{A. P. M. Place, L. V. H. Rodgers, P. Mundada, B. M. Smitham, M. Fitzpatrick, Z. Leng, A. Premkumar, J. Bryon, A. Vrajitoarea, S. Sussman, G. Cheng, T. Madhavan, H. K. Babla, X. H. Le, Y. Gang, B. Jäck, A. Gyenis, N. Yao, R. J. Cava, N. P. de Leon, and A. A. Houck, New material platform for superconducting transmon qubits with coherence times exceeding 0.3 milliseconds, Nat. Commun. \textbf{12}, 1779 (2021).}

\end{thebibliography}
\end{document}